 \documentclass[12pt,preprint]{aastex}

 \newcommand{\kms}{km\,s$^{-1}$}

 \shorttitle{What is the deuterium Abundance?}
 \shortauthors{Linsky et al.}

\usepackage{natbib}
\received{2005 May 24}
 
\begin{document}

 \title{WHAT IS THE TOTAL DEUTERIUM ABUNDANCE IN THE LOCAL GALACTIC  
DISK?\footnote{Based on observations made with the NASA-CNES-CSA {\it Far 
Ultraviolet Spectroscopic Explorer. FUSE} is operated for NASA by the Johns 
Hopkins University under NASA contract NAS5-32985.}}


 \author{Jeffrey L. Linsky\altaffilmark{2},
Bruce T. Draine\altaffilmark{3},
H. W. Moos\altaffilmark{4},
Edward B. Jenkins\altaffilmark{5},
Brian E. Wood\altaffilmark{2},
Cristina Oliveira\altaffilmark{4},
William P. Blair\altaffilmark{4},
Scott D. Friedman\altaffilmark{6},
Cecile Gry\altaffilmark{12},
David Knauth\altaffilmark{7},
Jeffrey W. Kruk\altaffilmark{4},
Sylvestre Lacour\altaffilmark{13},
Nicolas Lehner\altaffilmark{8},
Seth Redfield\altaffilmark{9},
J. Michael Shull\altaffilmark{10},
George Sonneborn\altaffilmark{11},
Gerard M. Williger\altaffilmark{4}}

\altaffiltext{2}{JILA, University of Colorado and National Institute of 
Standards and Technology, Boulder, CO 
80309-0440; jlinsky@jila.colorado.edu.}
\altaffiltext{3}{Department of Astrophysical Sciences, Princeton
University, Princeton, NJ 08544-1001.}
\altaffiltext{4}{Department of Physics and Astronomy, Johns Hopkins 
University, Baltimore, MD 21218.}
\altaffiltext{5}{Princeton University Observatory, Princeton, NJ 08544.}
\altaffiltext{6}{Space Telescope Science Institute, 3700 San 
Martin Drive, Baltimore, MD 21218.}
\altaffiltext{7}{Department of Physics and Astronomy, Northwestern 
University, Evanston, IL 60208-2900.}
\altaffiltext{8}{Department of Astronomy, University of 
Wisconson, 475 N. Charter St., Madison, WI 53706-1582.}
\altaffiltext{9}{University of Texas at Austin, McDonald 
Observatory, Austin, TX 78712. Harlan J. Smith Postdoctoral Fellow.}
\altaffiltext{10}{Center for Astrophysics and Space Astronomy, 
Dept. of Astrophysical and Planetary Sciences,  
University of Colorado, Boulder, CO 80309-0389.}
\altaffiltext{11}{Laboratory for Observational Cosmology, 
NASA Goddard Space Flight Center, Code 665, Greenbelt, MD 20771.}
\altaffiltext{12}{Laboratoire d'Astrophysique de Marseille, 13376 Marseille,
France.}
\altaffiltext{13}{Observatoire de Paris-Meudon, LESIA, 92195 Meudon, France.}


 \begin{abstract}

     Analyses of spectra obtained with the {\it Far Ultraviolet Spectroscopic
Explorer (FUSE)} satellite, together with spectra from the {\it Copernicus} 
and {\it IMAPS} instruments, reveal an unexplained very wide range 
in the observed deuterium/hydrogen (D/H) ratios
for interstellar gas in the Galactic disk beyond the Local Bubble.
We argue that spatial
variations in the depletion of deuterium onto dust grains can explain these
local variations in the observed 
gas-phase D/H ratios. We present a variable deuterium depletion model 
that naturally
explains the constant measured values of D/H inside the Local Bubble, 
the wide range of gas-phase D/H ratios observed in the
intermediate regime ($\log N({\rm H~I}) =$ 19.2--20.7), 
and the low gas-phase D/H
ratios observed at larger hydrogen column densities.  
We consider empirical tests of the deuterium depletion hypothesis:
(i) correlations of gas-phase D/H ratios with depletions of the refractory
metals iron and silicon, and
(ii) correlation with the H$_2$ rotational temperature. 
Both of these tests are consistent
with deuterium depletion from the gas phase in cold, not recently shocked, 
regions of 
the ISM, and high gas-phase D/H ratios in gas that has been shocked or 
otherwise heated recently. We argue that the most representative value for the
total (gas plus dust) D/H ratio within 1 kpc of the Sun is $\geq 23.1\pm 2.4$
(1~$\sigma$) parts per million (ppm). This ratio constrains Galactic chemical
evolution models to have a very small deuterium astration factor, the ratio
of primordial to total (D/H) ratio in the local region of the Galactic disk,
which we estimate to be $f_d \leq 1.19^{+0.16}_{-0.15}$ ($1\sigma$) or 
$\leq 1.12\pm 0.14$ ($1\sigma$) 
depending on the adopted light element nuclear reaction rates.

 \end{abstract}





\section{INTRODUCTION}

     Accurate measurements of the deuterium/hydrogen (D/H) ratio (by number)
provide critically important tests for models of primordial nucleosynthesis,
Galactic chemical evolution, and the chemical properties of the intergalactic
medium (IGM). Among all of the light elements, deuterium arguably provides 
the most
stringent tests of our understanding of these three critically important
topics, because the evolution of deuterium is predicted to be rather simple
and the dependence of D/H varies steeply with baryonic density.
\citet{eps76} and others have argued that
deuterium was formed only in the Big Bang and that D is easily converted 
to $^3$He, $^4$He, and heavier elements by nuclear reactions in stars, a 
process commonly called
astration. Thus, as stars evolve, the interstellar medium (ISM) receives
deuterium-depleted and metal-enhanced
material from stellar winds and supernova explosions, leading to a monotonic 
decrease in the D/H ratio with increased processing through stars.
Other light elements (e.g., $^3$He, $^4$He, $^7$Li) have more complex creation
and destruction histories or otherwise provide less sensitive tests of these
theories \citep[cf.][]{boe85,wil94}. Infall of less astrated gas, which is 
deuterium rich and metal poor, from the IGM and small galaxies 
captured by the Milky Way Galaxy, complicates models of Galactic chemical 
evolution, because the rate of infall, its time and spatial dependence,
and the timescale for mixing with other gas are poorly known.

     Although based on different physical processes and relating to very 
different times in the early universe, the two main approaches for 
determining the primordial D/H ratio, (D/H)$_{\rm prim}$, are now in 
agreement. The first approach is to measure the column densities of D~I and 
H~I in quasar absorption line systems.
\citet{kir03} have determined a mean value
of  D/H = $27.8^{+4.4}_{-3.8}$ parts per million (ppm) 
by averaging the log
(D/H) values using column densities $N$(H~I) and $N$(D~I) obtained from 
absorption line measurements of gas in the
lines of sight toward five quasars. Most of these measurements use high 
resolution spectra obtained with the {\it Keck} telescope.
Since the observed gas has very low metallicity, these values of D/H 
are likely close to, but perhaps slightly
smaller than, the primordial (Big Bang) value, (D/H)$_{\rm prim}$. 
The measurement of D/H = $22\pm 7$ ppm for the high velocity
cloud Complex C \citep{sem04} likely refers to gas located 6--10 kpc 
into the Galactic halo
with low metallicity, 10--30\% solar abundance \citep{col03},
and this value should be somewhat smaller than (D/H)$_{\rm prim}$.

     The second approach involves the analysis of 
data from the {\it Wilkinson Microwave Anisotropy Probe (WMAP)}
satellite and other cosmic microwave background experiments.
Assuming a flat $\Lambda$-dominated universe and other standard assumptions,
several authors \citep{spe03,cyb03,coc04,san06} 
have derived tight constraints
on the baryon closure parameter, $\Omega_{\rm b}$, and the Hubble constant.
Together with other constraints from the power spectrum of galaxy clustering
and L$\alpha$ forest data, they find that $\Omega_b h^2 = 0.0224$ with a 
1$\sigma$ error of about 4\%. These results are summarized in Table~1.
The ratio of baryons to photons (multiplied by $10^{10}$), $\eta_{10}$, is 
simply related to $\Omega_b h^2$ \citep{bur01}. In the standard
big bang nucleosynthesis (SBBN) model, $\eta_{10}$ is the single parameter that
predicts the primordial light element abundances including (D/H)$_{prim}$. 
For example, \citet{spe03}, \citet{coc04}, and \citet{san06} obtain 
(D/H)$_{\rm prim} \approx 26$~ppm. Using the same values of $\Omega_b h^2$ 
and $\eta_{10}$ but a more recent compilation of nuclear reaction rates 
\citep{ang99}, \citet{cyb03} obtain    
(D/H)$_{\rm prim} = 27.5^{+2.4}_{-1.9}$ ppm. 
The somewhat different values of (D/H)$_{\rm prim}$ arise from the different 
authors treating the light element nuclear reaction rates and error 
propagation somewhat differently (cf. Nollett \& Burles 2000).
We therefore use in the subsequent analysis two different values for 
(D/H)$_{\rm prim}$, $27.5^{+2.4}_{-1.9}$ ppm \citep{cyb03} and 
$26.0^{+1.9}_{-1.7}$~ppm \citep{coc04}. Both values are consistent with the
quasar absorption line result for (D/H)$_{\rm prim}$.

     Beginning with observations by the {\it Copernicus} satellite, data from 
high-resolution ultraviolet spectrometers on several satellites
have been analyzed to measure D/H in the Galactic ISM.
These results provide insight into the abundance of deuterium in the present 
epoch, which with measurements of various metal abundances constrain
models of Galactic chemical evolution. The {\it International Ultraviolet 
Explorer (IUE)} satellite and more recently the {\it Goddard High Resolution 
Spectrometer (GHRS)} and {\it Space Telescope Imaging Spectrograph (STIS)} 
instruments on the {\it Hubble Space Telescope (HST)} have provided UV 
spectra of the Ly$\alpha$
absorption lines of H~I (1215.67 \AA) and D~I (shifted by --82 km s$^{-1}$
relative to the H~I line), which are needed to measure D/H in the ISM along
lines of sight to stars located within a few hundred parsecs of the Sun. The
same analysis technique has been applied to hydrogen and deuterium lines
higher in the Lyman series
that lie in the spectral range 912--1025 \AA, using data obtained with the
{\it Copernicus}, the {\it Interstellar Medium Absorption Profile Spectrograph 
(IMAPS)}, and the {\it FUSE} satellite. 
For a description of the {\it FUSE} satellite and its operations, see
\citet{moo00} and \citet{sah00}.

The far-UV
spectral range is essential for sampling more distant lines of sight using hot
subdwarfs and white dwarfs as background sources, because the decreasing
opacity in the higher Lyman lines allows one to observe the 
deuterium lines without masking by the nearby very opaque broad hydrogen lines.
For example, the dark core of the H~I Ly$\alpha$ line absorption broadens
to mask the D~I Ly$\alpha$ line when the hydrogen column density 
$\log N({\rm H~I}) > 18.7$, but at Ly$\gamma$ (972~\AA) 
the horizon for detecting the
DI Ly$\gamma$ line extends to $\log N({\rm H~I}) \approx 20.0$. Measurements of
D/H using the higher Lyman series lines are feasible to about 
$\log N({\rm H~I}) \approx 21$ (cf. Hoopes et al. 2003), 
where the density of overlapping H$_2$ lines and the large widths
of the interstellar H~I Lyman lines determine the final horizon for such
analyses.

     There are several subtle effects that can enter the analysis of 
D~I and H~I 
lines in the far UV. One disadvantage is that, for long lines of sight, 
there is 
an increased chance of weak features at large velocity displacements. 
An H~I feature near --82 km s$^{-1}$ can masquerade as D~I absorption. 
An important advantage is that the far UV includes a large number of 
Lyman lines with a wide range of optical depths. Thus, measurements of 
$N$(H~I) and $N$(D~I) can be based on several spectral lines in the 
Lyman series. Another advantage is that, for the higher Lyman lines, 
the upper state lifetimes 
are long, decreasing the damping and better separating the H~I and D~I 
absorption for each line in the Lyman series and between adjacent lines in the
series.

     Since the very first of these measurements, a disturbing aspect of 
the D/H studies has been the wide range of measured values with no apparent 
pattern in the data or a convincing explanation for it. 
We address the first part of this problem with the
latest data, including many measurements with {\it FUSE}; we conclude 
(see \S~2) that the variations are real, and we show that there is a 
consistent pattern. We propose that
the variations arise from time-dependent deuterium depletion 
onto dust grains (see \S~4 and \S~5), and we examine two
critical tests (see \S~6) of this model. The first test (\S6.1) uses iron and 
silicon depletions as proxies for grain destruction, and the second test 
(\S6.2) uses correlations of D/H 
with H$_2$ excitation temperatures. We conclude (see \S~7 and \S~9) 
that the most likely value for the total
D/H ratio within 1 kpc of the Sun is about 45\% larger than the previously 
assumed value of $\approx$15 ppm, which provides a revised constraint 
on models of Galactic chemical evolution. Also, the 
inferred depletion of deuterium from the gas phase places constraints
on the physics and chemistry of interstellar dust. See \citet{lin06} for 
an early version of this paper.

\section{MEASUREMENTS OF D/H IN THE LOCAL GALACTIC DISK}

     Using {\it FUSE} data, \citet{woo04} measured (D/H)$_{\rm gas}$ 
along the lines of sight to
two hot subdwarf stars, JL 9 and LS 1274, located at distances of $590\pm 160$
pc and $580\pm 100$ pc, respectively. They also summarized the 
(D/H)$_{\rm gas}$ measurements
observed by the previously mentioned satellites for 38 lines of sight to stars 
located between 3.2~pc and 2200~pc from the Sun. These column densities 
are summarized in Table~2 together with the new results for the  
PG~0038+199 \citep{wil05}, HD~90087 \citep{heb05}, LSE~44 \citep{fri06},
WD~1034+001 and TD1~32709 \citep{oli06} 
lines of sight and (D/H)$_{\rm gas}$ for Lan~23 \citep{oli03}, which was 
not included in previous compilations owing to the large uncertainty in 
(D/H)$_{\rm gas}$. 
Table~3 lists the (D/H)$_{\rm gas}$ ratios computed from 
the measured column densities in 
Table~2, and the 1$\sigma$ errors in (D/H)$_{\rm gas}$ 
are computed from the square root 
of the sum of the squares of the percent errors in $N$(D~I) and $N$(H~I).
When authors present their results
for (D/H)$_{\rm gas}$ but not $N$(D~I), we have computed $N$(D~I) and its 
uncertainty to be consistent with the published values of 
(D/H)$_{\rm gas}$ and $N$(H~I). 
The (D/H)$_{\rm gas}$ measurements plotted in
Figure~1 show a complex picture that, as we shall see, seriously challenges 
previously held ideas concerning Galactic chemical evolution. 

     Table 2 lists the most recent measurements of $N$(H~I) and $N$(D~I),
ordered by increasing $N$(H~I),
for interstellar gas along the lines of sight to the listed stars, 
obtained using the {\it FUSE, Copernicus, IMAPS}, and {\it GHRS} and 
{\it STIS} instruments on {\it HST}. (D/H)$_{\rm gas}$, which is derived from 
the $N({\rm D~I})/N({\rm H~I})$ ratio, is not always the same as 
originally reported because
there are measurements of $N$(H~I) obtained more recently with higher S/N
and/or spectral resolution than the corresponding $N$(D~I) measurements. 
This is typically the case when the 
only measurement of $N$(D~I) was obtained with the {\it Copernicus} 
satellite and
$N$(H~I) was later measured from {\it HST} spectra. For the sightlines to  
$\mu$~Col, $\beta$~CMa, $\theta$~Car, and $\lambda$~Sco, we list $N$(H~I) and
$N$(D~I) for a specified velocity component (see notes to Table~2)
rather than for the entire 
inhomogeneous line of sight, since the high quality spectra
enabled the authors to identify the velocity components and measure the 
column densities in each component, although the components were likely 
not resolved.

For the sightlines
that extend through the Local Bubble to more distant stars, we also list
(D/H)$_{\rm gas-LB}$, the value of (D/H)$_{\rm gas}$ for the line of sight 
beyond the Local Bubble foreground, 
obtained by subtracting $N({\rm H~I})_{\rm LB}=10^{19.20}$ cm$^{-2}$ and
$N({\rm D~I})_{\rm LB}=10^{14.39}$ cm$^{-2}$ from the measured values of
$N$(H~I) and $N$(D~I) for the full line of sight.  
We assume that the 
Local Bubble extends to $\log N({\rm H~I}) = 19.2$, 
since beyond that point the 
(D/H)$_{\rm gas}$ values are no longer consistent with a constant value. 
\citet{sfe99} place the edge of the Local Bubble at 
$\log N({\rm H~I})\sim 19.3$, which corresponds 
to a 20 m\AA\ equivalent width of the Na~I 5892~\AA\ line formed in 
cold interstellar gas. Within
the Local Bubble, we adopt (D/H)$_{\rm gas} = 15.6\pm 0.4$~ppm \citep{woo04}. 
Given the complex three dimensional shape of 
the Local Bubble \citep{lal03}, these assumptions for its 
column densities are approximations, but 
they provide good first estimates of the foreground required 
for isolating the deuterium and hydrogen column 
densities of the gas lying beyond the Local Bubble. This is especially 
important for 
lines of sight that do not extend very far beyond the Local Bubble, 
but this correction is small for 
the more distant lines of sight because the Local Bubble 
contribution to the total is minor. Those lines of sight that
are definitely inside the Local Bubble are flagged as LBg in the eighth column
of Table~2. 

     Figure 1 shows the (D/H)$_{\rm gas}$ ratios for the 47 lines of sight as
functions of hydrogen column density. The (D/H)$_{\rm gas}$ ratios extend over 
an unexpectedly wide range, but \citet{woo04} proposed that these results
naturally fit into three regimes. For $\log N({\rm H~I}) < 19.2$ (in cm$^{-2}$
units), corresponding to sightlines extending to the edge of the Local Bubble
near 100 pc, (D/H)$_{\rm gas}$ has a constant value of $15.6\pm 0.4$ ppm, 
where the
quoted uncertainty is the 1$\sigma$ error in the mean for the 23 lines of sight
within this regime. This constant value for the local (D/H)$_{\rm gas}$ ratio 
had been previously established based on {\it HST} data \citep{lin98,lin03}, 
but the
addition of new lines of sight observed by {\it FUSE} \citep{moo02} strongly
supports this result. The very different values of 
(D/H)$_{\rm gas}$ measured in the distant ($\log N({\rm H~I}) > 20.7$) and 
intermediate ($\log N({\rm H~I}) = 19.2 - 20.7$) regimes beyond the Local 
Bubble are described at the end of this section.

     \citet{heb03}, \citet{heb05} and others have derived (D/H)$_{\rm gas}$ 
in an alternative way from the product 
(D/O)$_{\rm gas}\times$ (O/H)$_{\rm gas}$. Since the 
charge-exchange reaction between O~I and H~I is fast, 
the ionization of oxygen, hydrogen, and deuterium are very nearly the same.
In the ionization equilibrium calculations for 
their best models (Models 2 and 8) for the 
Local Interstellar Cloud, \citet{fri03} find that each element 
is about 30\% ionized.
Thus in the gas phase (O/H) = $N$(O~I)/$N$(H~I) = 
[$N$(O~I)+$N$(O~II)]/[$N$(H~I)+$N$(H~II], and a similar relation can be 
written for (D/O). Empirically, the abundance
of O relative to H shows little evidence for spatial variation
\citep{mey98,and03,car04}. Also, the outcomes for (D/O)$_{\rm gas}$ may not
suffer from the effects of line saturation, as there are O~I lines that
are optically thin. All three of these considerations should make
O~I a good proxy for H~I. The (D/O)$_{\rm gas}$ ratio should therefore 
behave in a very similar way to (D/H)$_{\rm gas}$. 
The problems with this technique are that uncertainties in two different 
measurements must be considered instead of one, and the amount of depletion
of oxygen from the gas phase may depend on 
density \citep{car04}.  Table~4 summarizes the
recent measurements of (O/H)$_{\rm gas}$. \citet{oli05} found that inside the 
Local Bubble, (O/H)$_{\rm gas}$ is constant with a mean value of 
$345\pm 19$~ppm.
\citet{mey98}, \citet{mey01}, and \citet{and03} find slightly higher values of
(O/H)$_{\rm gas}$ for lines of sight extending beyond the Local Bubble with 
perhaps a slight decrease in (O/H)$_{\rm gas}$ at large column densities 
(log N(H~I)$>21.5$). 
We assume (D/O)$_{\rm gas} = (3.84\pm 0.16)\times 10^{-2}$ ($1\sigma$), 
the mean value 
that \citet{heb03} find for lines of sight inside the Local Bubble, and compute
the values of (D/H)$_{\rm gas}$ listed in Table~4. Inside the Local Bubble, 
using the value of (O/H)$_{\rm gas}$ obtained by \citet{oli05}, we compute 
(D/H)$_{\rm gas} = 13.2\pm 0.9$~ppm. For values of (O/H)$_{\rm gas}$ derived 
from longer lines of sight \citep{and03}, 
we find (D/H)$_{\rm gas} = 13.9\pm 0.7$ to 
$15.7\pm 0.9$~ppm, depending on the number of lines of sight included. 
In their Models 2 and 8, \citet{fri03} conclude that (O/H)$_{\rm gas} = 
380\pm 30$\footnote{We arbitrarily assign an error of 8\% to their value of
(O/H)$_{\rm gas}$ based on the discussion in their paper.}~ppm,
leading to (D/H)$_{\rm gas} = 14.6\pm 1.3$~ppm. 
These alternative ways of determining (D/H)$_{\rm gas}$ are in agreement 
or smaller by 1--2$\sigma$ compared with 
the direct measurement of (D/H)$_{\rm gas} = 15.6\pm 0.4$~ppm obtained by 
\citet{woo04}. 

     To put these
numbers into perspective, we note the recent measurements of the solar ratio,
(O/H)$_{\odot} = 458\pm 53$~ppm \citep{asp04} and $390\pm 63$~ppm 
\citep{mel04}. \citet{bah05} argue that the \citet{asp04} value 
for the solar oxygen abundance is too low to explain the helioseismological
determinations of convection zone parameters and surface helium composition,
but a recent increase in the estimated opacity near the base of the solar 
convective zone \citep{sea04} minimizes the problem even for the 
\citet{mel04} low abundance. 
In \S3 we argue that about 120~ppm of oxygen could be 
depleted on to grains if Mg, Fe, and Si have solar abundances in the ISM, 
but \cite{fri03} conclude that the ISM abundances of these elements
may be only 60--70\% solar.
We therefore take the measured solar value ($458\pm 53$ ppm) 
and that value 
less 120~ppm as a plausible range in (O/H)$_{\rm gas}$. Then the plausible 
range in the inferred
(D/H)$_{\rm gas}$ = (D/O)$_{\rm gas}\times$ (O/H) 
is $13.0\pm 2.0$~ppm to $17.6\pm 2.1$~ppm. All of the Local 
Bubble values of (D/H)$_{\rm gas}$ measured directly and indirectly 
(see Table~4) fall well 
within this range. The differences among these values and with the 
directly measured (D/H)$_{\rm gas} = 15.6\pm 0.4$~ppm
may represent measurement uncertainty
and perhaps some systematic effects. 

     An important new result noted for (D/H)$_{\rm gas}$ measurements by 
\citet{heb03} and emphasized by \citet{woo04}
is that sightlines with large
$N$(H~I) appear to have systematically low values of (D/H)$_{\rm gas}$.
\citet{woo04} called attention to the clump of four
low (D/H)$_{\rm gas}$ values at $\log N({\rm H~I}) > 20.7$, which we refer to 
as the distant regime. Whereas 
unambiguously low values for D/H in
lines of sight extending beyond the Local Bubble have been known since the
{\it Copernicus} results for $\delta$~Ori, $\epsilon$~Ori, and $\theta$~Car 
\citep{jen99,lau79,all92},
the new results for JL 9 and LSS 1274, together with the
(D/H)$_{\rm gas}$ values determined from {\it FUSE} observations of HD 195965 
($794\pm 200$ pc) and HD 191877 ($2200\pm 550$ pc) from \citet{hoo03} 
and HD~90087 ($2740\pm 800$~pc) from \citet{heb05} indicate an
emerging pattern. For these five long lines of sight, 
(D/H)$_{\rm gas}$ = $8.6\pm 0.8$ (standard deviation of the mean) ppm, 
nearly a factor of two below the Local Bubble value. 

     The intermediate regime for $\log N(H~I) =$ 19.2--20.7 exhibits a factor
of 4--5 range in (D/H)$_{\rm gas}$: lines of sight to five stars 
have high values (near 22 ppm), five
stars have low values (near 6 ppm), and nine stars have intermediate values.
Since (i) the {\it Copernicus, IMAPS} and {\it FUSE} data all show this wide 
spread in (D/H)$_{\rm gas}$ values in the intermediate regime, 
(ii) there does not appear to
be any correlation of (D/H)$_{\rm gas}$ with Galactic latitude, longitude, or
distance, and (iii) the (D/O)$_{\rm gas}$ values also show a large spread,
we conclude
that the wide spread in (D/H)$_{\rm gas}$ in the intermediate regime is not 
due to instrumental artifacts or unknown systematic errors, but is rather a 
robust result that requires a sensible scientific explanation. 
We call attention to the five high data points in the intermediate regime 
with (D/H)$_{\rm gas}\approx 22$ that will play an important role in our
deuterium depletion model (see \S 5).

\section{PREVIOUS EXPLANATIONS OF THE D/H DATA}

     The constant value of (D/H)$_{\rm gas}$ inside the Local Bubble suggests 
that this matter has experienced a similar history of nuclear
processing, mixing, and deuterium depletion (cf. \S~5). 
This history consists of star formation in the
Scorpio-Centaurus Association some 10 Myr ago and the creation of the Local
Bubble by supernovae in the Lower Centaurus Crux subgroup of the
Scorpio-Centaurus Association \citep{mai01,ber02}. Outside the Local Bubble,
interstellar gas could have experienced 
different chemical evolution resulting from
different rates of astration, infall of nearly primordial material from the
Galactic halo or the IGM, and insufficient mixing 
to homogenize the Galactic ISM. However, Galactic chemical evolution models are
faced with the severe challenge of explaining the factor of 4--5 range in
(D/H)$_{\rm gas}$ ratios just outside of the Local Bubble and the clumping of 
low (D/H)$_{\rm gas}$ values for the lines of sight with the largest values of 
$N$(H~I). Recent Galactic
chemical evolution models constructed to fit a wide range of constraints,
including the abundances of deuterium, $^3$He, $^4$He, Li, and heavier
elements created by astration, generally predict a moderate astration
factor for the total amount of (D/H)$_{\rm total}$ in all forms, f$_d =$
(D/H)$_{\rm prim}$/(D/H)$_{\rm total} \approx 1.5$ \citep[cf.][]{chi02,rom03}. 
If we assume (D/H)$_{\rm prim} = 27.5$~ppm, then the astration factors 
in the intermediate
regime mentioned in the previous section would lie in the range of 1.25 to 4.6.
How could such a large range in astration factors occur over distance scales of
only 100--500 pc, but an apparently constant astration factor of 3.2 appear to
be representative for longer distance scales? 

     Galactic chemical evolution models generally include enrichment of gas in
the Galactic disk by the infall of less-astrated material from the IGM or gas
from neighboring small galaxies. For example, \citet{gei02} find that the Local
Bubble value of (D/H)$_{\rm gas}$ could be explained by a mixture of 60\% gas
representative of the protosolar nebula (with additional chemical processing
over the last 4.6 Gyr) and 40\% infalling gas with high (D/H)$_{\rm gas}$ and 
low
metals representative of the Large Magellanic Cloud. While this mechanism may
explain the Local Bubble abundances, variations of the percentages to explain
the wide range of observed (D/H)$_{\rm gas}$ values would require that the
infalling gas be spatially concentrated into small regions of the Galactic disk
and then not be mixed. We note here that typical sizes for structures in 
the nearby infalling high-velocity cloud Complex C are about 100 pc (Collins,
Shull, \& Giroux 2003). We discuss the infall hypothesis further in \S~8.

     Both spatially variable astration and infall models may, in principle,
explain the (D/H)$_{\rm gas}$ measurements, but they require extreme 
assumptions concerning the deuterium destruction and infall rates 
to explain the observed (D/H)$_{\rm gas}$ ratios. 
Simulations with their three-dimensional supernova-driven ISM model 
led \citet{dea02} to conclude that for the Galactic 
supernova rate, it takes some 350 Myr to completely mix the ISM on distance 
scales from parsecs to kiloparsecs. 
In view of this relatively rapid mixing rate, we expect the total 
interstellar D/H ratio in all forms, (D/H)$_{\rm total}$, to show little 
variation on length scales less than 1 kpc, unless one invokes localized 
starbursts sufficiently 
strong for astration to produce significantly lower (D/H)$_{\rm total}$ on a
timescale shorter than 350 Myr. 
Both the variable astration and infall models fail to explain the absence
of any correlation between the observed (D/H)$_{\rm gas}$ and (O/H)$_{\rm gas}$
ratios (e.g., Draine 2005), which should show a negative correlation for
both models. We explore instead another physical process
that could have a much larger effect on local values of the (D/H)$_{\rm gas}$ 
ratio than either astration or infall.

Figure \ref{fig:DvsO} shows observed values of (D/H)$_{\rm gas}$ and 
(O/H)$_{\rm gas}$ for all sightlines that have
published column densities of H, D, and O with $1\sigma$
uncertainties of less than 0.1 dex for each.
The error bars in the figure are $1\sigma$; because 
$N({\rm H~I + 2H_2)}$ appears in the denominator of both
abscissa and ordinate, errors in (O/H) and (D/H) are partially
correlated.
Variations in astration or infall would produce anticorrelation of 
(D/H)$_{\rm total}$ and (O/H)$_{\rm total}$.
Figure \ref{fig:DvsO} shows the expected track of (D/H)$_{\rm total}$ 
vs. (O/H)$_{\rm gas}$
for chemical evolution model C-I of Chiappini et al.\ (2002),
where (i) we have taken Chiappini et al.'s variation with galactocentric
radius as a proxy for variations in astration and infall, (ii) we
have rescaled the predicted (D/H)$_{\rm total}$ values slightly 
to be consistent with primordial (D/H)$_{\rm total}\approx$27.5~ppm 
estimated from quasar absorption lines and {\it WMAP}, 
and (iii) we assume partial depletion of O into grains (see below), with
(O/H)$_{\rm gas}=0.75$(O/H)$_{\rm total}$.
The observed (D/H)$_{\rm gas}$ values fall systematically below the
(D/H)$_{\rm total}$ values expected for at least this astration/infall model.
The full (factor of $\sim$3) range of variation in (D/H)$_{\rm gas}$
is seen among the 6 sightlines with (O/H)$_{\rm gas}$ between
300 and 390~ppm.
Variable astration and infall may account for a fraction of the
variation in (D/H)$_{\rm gas}$, but it appears that 
most of the variation must be
due to some other mechanism --- such as depletion of D onto dust grains.

The median value of (O/H)$_{\rm gas}$ in Figure~2 is 363~ppm, 
but it is likely that another
$\sim 120$~ppm is locked up in grains in silicates or oxides
(assuming the grains contain $\sim$90\% of solar abundances of Mg, Fe, and Si,
with composition Mg$_x$Fe$_{2-x}$SiO$_4$).\footnote{
   The abundance of H$_2$O is negligible outside of dark clouds.}
From these well-characterized sightlines, we therefore
estimate (O/H)$_{\rm total}\approx 483$ ppm in the local ISM,
slightly higher than but consistent with the recent estimate 
for the solar abundance
(O/H)$_\odot=458\pm53$~ppm (Asplund et al.\ 2004).
Thus we estimate that O$_{\rm gas}/$O$_{\rm total}\approx(363/483)=0.75$ for
diffuse regions.
Note that for a primordial D/H = 27~ppm 
and (O/H)$_{\rm gas}/$(O/H)$_{\rm total}\approx0.75$,
the Chiappini et al. astration + infall model C--I (see Figure~2)
has (D/H)$_{\rm total}$ 
reduced to 23 ppm by the time (O/H)$_{\rm gas}$ has risen to 
$\approx 360$~ppm, very close to the measured value of $345\pm 19$~ppm
\citep{oli05} for the solar neighborhood.
We will return to this point in \S~8.

\section{DEUTERIUM DEPLETION ONTO DUST GRAINS}

     It is important to recognize that all of the direct (D/H)$_{\rm gas}$
measurements refer to interstellar gas, whereas the Galactic chemical evolution
and primordial nucleosynthesis models refer to (D/H) in all forms,
(D/H)$_{\rm total}$. Dust
grains exist in both the cold (10--$10^2$~K) and warm 
($10^3$--$10^4$~K) phases of the ISM.
In diffuse interstellar clouds, 50--90\% of elements such as
Mg, Si, and Fe reside in grains
\citep[e.g.,][]{sem96,jen04}, 
and gas-phase abundances of such elements relative
to hydrogen are much lower than the total elemental abundances. A similar
process may preferentially deplete deuterium in the gas phase relative to
hydrogen. This could lead to large spatial variations in the gas-phase ratio,
(D/H)$_{\rm gas}$, while the total deuterium abundance, (D/H)$_{\rm total}$,
including both gas and  dust, remains roughly constant. If deuterium
depletion onto dust grains is important, then the total Galactic D/H 
ratio cannot be extracted from gas-phase measurements that ignore deuterium 
locked in dust grains.

     \citet{jur82} first proposed that the depletion of deuterium onto dust
grains might explain the factor of two difference in (D/H)$_{\rm gas}$ values 
for the
lines of sight to $\zeta$~Pup and $\delta$~Ori obtained from the analysis of
{\it Copernicus} spectra \citep{vid77,lau79}. Jura argued that the sticking
probability of deuterium onto dust grains should be high and that deuterium
should be more tightly bound to the dust than hydrogen, since the zero-point
energies of deuterium-metal bonds are lower than for the
corresponding hydrogen-metal bonds. Numerical calculations of the chemistry of 
grain mantles by Tielens (1983) showed that deuterium can be highly 
enriched in grains, because the lower zero-point energy for deuterium bonds 
as compared to corresponding hydrogen bonds strongly favors the 
formation of deuterated molecules 
like HDCO on grain surfaces (cf. Turner 1990), effectively removing deuterium 
from the gas phase in dark regions where grains are coated with ices.

\citet{dra04,dra06} developed this idea further by noting that the C--D 
bond energy in polycyclic aromatic hydrocarbon (PAH) molecules
is 0.083~eV larger than the C--H bond.
If H and D in the gas were in thermodynamic equilibrium with
deuterated PAH material, one would have 
(D/H)$_{\rm dust}$/(D/H)$_{\rm gas}\approx e^{970K/T_{\rm dust}}$, 
which exceeds $5\times10^4$
for $T_{\rm dust} < 90$~K.  Although thermodynamic equilibrium arguments
must be regarded with suspicion in the nonequilibrium ISM, Draine
argued that it is plausible to consider that in a
steady-state the grains might have (D/H)$_{\rm dust}$/(D/H)$_{\rm gas}$ as
high as $5\times10^4$ or more.  

Contemporary grain models require a substantial amount of hydrogen-bearing
carbonaceous material.
From the relative strengths and profiles of different C-H stretching modes,
\citet{pen02} estimate that $\sim$85\% of the  
carbon in dust is in aromatic form, 
with H/C$\approx$0.35, 
and $\sim$15\% of the carbon is aliphatic, with H/C$\approx$2.1.
If (C in grains)/(H total) $\approx200$~ppm, as estimated by \citet{dra04},
then the aromatic hydrocarbon material
would contain 
(H in aromatics)/(H total) $\approx60$~ppm.
Suppose that in the local Galactic disk
(D/H)$_{\rm total}=22$~ppm, and that in the aromatic material 
the D/H ratio is enhanced so that $({\rm D/H})_{\rm dust}=0.27$.
Then ${\rm (D/H)}_{\rm gas} = {\rm (D/H)}_{\rm total} - 
{\rm (D/H)}_{\rm dust}\times {\rm (H)}_{\rm dust}/{\rm H}_{\rm total}
\approx 6{\rm\,ppm}$.
Therefore we see that if D enrichment can bring the aromatic grains
to (D/H)$_{\rm dust}\approx0.27$, the gas phase deuterium abundance
could be reduced to $\sim 6$~ppm, which would be sufficient
to explain the lowest values of (D/H)$_{\rm gas}$ observed.
This would require extreme enrichment of the grains, relative to the gas, by
a factor 
(D/H)$_{\rm dust}$/(D/H)$_{\rm gas}\approx 0.27/6{\rm\,ppm}\approx5\times10^4$.
This is possible for $T_{\rm dust} < 90$~K (see above).
If the aliphatic and aromatic hydrocarbons are equally D-enriched, then
(D/H)$_{\rm dust}/{\rm (D/H)}_{\rm gas}\approx2\times10^4$ would suffice to
explain the sightlines with the lowest values of
(D/H)$_{\rm gas}\approx6$~ppm, with
(D/H)$_{\rm dust}\approx0.13$.

     Figure 1 shows that (D/H)$_{\rm gas}$ varies from one
sightline to another, covering a range from $\sim$6~ppm to $\sim$22~ppm.
\citet{dra04, dra06} and \citet{woo04} suggest a dynamic model 
of the ISM in
which deuterium is depleted from the gas phase onto dust grains over time until
the D-bearing grains are eroded or destroyed by sputtering or 
grain-grain collisions
in strong shocks from supernova remnants, or by strong UV
radiation fields from nearby hot stars, thereby returning
the sequestered deuterium back to the gas phase. Is this idea supported by
observations, and can it explain the three regime interpretation for the
(D/H)$_{\rm gas}$ data shown in Fig.~1? 

     The most direct test of the deuterium depletion model would be to measure
(D/H)$_{\rm dust}$ in interstellar dust grains. \citet{kel00} have performed 
this
critical experiment by measuring the deuterium abundance in interplanetary dust
particles containing amorphous carbonaceous material. They argue that sample
L2009*E2, which was captured in the upper atmosphere of the Earth, originated
in the ISM because of the very high D/H ratios measured for various molecules
(which indicates formation at low temperatures where fractionation is
important),
the elevated $^{15}$N/$^{14}$N ratio, and the particle size distribution. For
the whole dust grain sample, they find that (D/H)$_{\rm dust} = 220$ ppm, 
but for one fragment of the sample (D/H)$_{\rm dust} = 1900$ ppm. 
If we assume that the
local ISM value of (D/H)$_{\rm gas}$ is 15 ppm, then the (D/H)$_{\rm dust}$ 
enhancement
factors are 15 for the entire sample and 130 for the fragment. This
measurement alone provides a proof of concept that deuterium depletion onto
grains does occur in the ISM, although the highest D/H ratio measured by 
\citet{kel00} in this particle sample falls a factor of 
$\sim70$ short of the factor of $\sim10^4$ that is required for grain
deuteration to explain the typical
observed variations in (D/H)$_{\rm gas}$.
It is interesting that high values of
(D/H)$_{\rm dust}$ are only detected in amorphous organic (i.e.,
carbonaceous) material in extraterrestrial samples \citep{kel00, pee04}. 

     As part of their paper summarizing the observational evidence for 
deuterated PAHs (hereafter called PADs) in the ISM,
\citet{pee04} point out that deuterium fractionations (the ratio of deuterated
molecules to undeuterated molecules) for simple molecular species are 
typically 0.01 to 0.1 (e.g., Markwick, Charnley \& Millar 2001) and
higher deuterium fractionations have been measured (e.g., Parise et al. 2002).
Even triply deuterated ammonia \citep{tak02}
and methanol \citep{par04} have even been observed in cold
interstellar clouds.
\citet{pee04} then argue that PADs are more stable, more abundant in the
ISM compared to many of the simpler molecules that show deuterium enrichments,
and carry more H atoms per molecule. PADs may, therefore, represent a large
reservoir of deuterium-enriched species.

     \citet{pee04} then call attention to
infrared emission features at 4.4~$\mu$m and 4.65~$\mu$m in ISO spectra of 
the Orion Bar and M17 and identify these features as C--D stretching modes 
in PADs. From the relative strengths of the emission bands of PADs and PAHs,
\citet{pee04} estimate that the D/H ratio in these molecules is
$0.17\pm 0.03$ for the Orion Bar and $0.36\pm 0.08$ for M17. These ratios are
about $10^4$ times larger than for atomic deuterium and hydrogen in the gas
phase, (D/H)$_{\rm gas}$. 
It is striking that the D/H ratios required to account for the
observed 4.4~$\mu$m and 4.65~$\mu$m emission are comparable to the
degree of deuteration of PAH material argued for by \citet{dra04, dra06},
and which imply substantial depletion of D from the gas if
this D/H ratio were characteristic of all of the hydrocarbon
material in grains.  Thus, deuteration of PAHs over time in
a cold environment could be an important process for removing deuterium from
the gas phase, and \citet{pee04} describe four  different astrochemical
processes by which this could occur. 


     Grains must be cold for the deuterium depletion process to operate on
their surfaces. Even in the 7000 K gas in warm interstellar clouds inside the
Local Bubble, the grains are cold 
because at the densities of diffuse clouds, grain heating is by
absorption of UV photons from stars rather than by collisions with the gas;
the grains cool by the emission of infrared radiation. The observed IR emission
spectrum for wavelengths greater than 80 microns requires that typical grain
temperatures are $\sim$18~K \citep{dra03}. 

Because (D/H)$_{\rm gas}$ varies from one sightline to another, D
depletion and return of D to the gas must be dynamic processes.
What is the timescale for D depletion?
\citet{dra06} considers reaction of D$^+$ ions with neutral PAHs, 
and reaction of D atoms with PAH$^+$ ions, arguing that the impinging D will
become chemically bound in the PAH, displacing a previously-bound H if
necessary.  
Considering the charge distribution of the PAH population under 
``cold neutral medium'' conditions ($n_{\rm H}\approx 30{\rm\,cm}^{-3}$,
$n_e\approx0.03{\rm\,cm}^{-3}$, $T\approx 10^2$~K),
\citet{dra06} estimates that the timescale for a D atom
to collide with a charged PAH is $\sim$2~Myr.
If no processes act to return D to the gas phase, D depletion by a factor of
$e$ could therefore be achieved in only 2~Myr. 
In more diffuse regions, the timescale will be longer.  For ``warm neutral
medium'' conditions ($n_{\rm H}\approx 0.3{\rm\,cm}^{-3}$,
$n_e\approx0.03{\rm\,cm}^{-3}$, $T\approx 6000$~K), a similar calculation
gives a much longer D depletion time of $\sim$50~Myr.  
Therefore it appears that
D depletion will be dominated by the cooler, denser regions where gas-grain
collision rates are high.

An important point is that most of the surface area available for
deuterium depletion is provided by the population of free-flying
PAHs, but this population is thought to contain only 
$C_{\rm PAH}/H_{\rm total}\approx 60$~ppm \citep{lid01}, or $\sim$1/3 
of the total aromatic material, with the remaining carbonaceous material
in grains with radii $a\gtrsim 150$~\AA.  How can the material in
the larger grains become D-enriched?
In denser regions, free-flying PAHs are expected to coagulate and become
part of larger grains.  The population of small PAHs is presumed to
be replenished by grain fragmentation in grain-grain collisions.
Thus there is steady interchange of material between PAHs and large grains.
The timescale for a PAH to collide and coagulate with larger grains
is $\sim$10~Myr in the cold neutral medium, assuming that the larger grains
have a total projected area per H nucleon of $10^{-21}{\rm\,cm}^2$, and
a characteristic velocity relative to the gas of $\sim1$~km/s, driven
by MHD turbulence \citep{yan04}.
In warm neutral medium conditions,  
this timescale becomes very long; once again, it appears that the exchange of
matter between PAHs and larger grains is dominated by processes in
the cold neutral medium. Therefore,
the conditions are favorable for the time-dependent depletion of deuterium from
the gas phase due to local events without changing (D/H)$_{\rm total}$ 
significantly by either astration or infall of less-astrated gas. 
  
\section{TIME-DEPENDENT DEUTERIUM DEPLETION MODEL}

     We now apply the deuterium depletion model to the three-regime
interpretation of the (D/H)$_{\rm gas}$ data presented 
in Figure~1. The measurements
of (D/H)$_{\rm gas}$ in the Local Bubble are for gas in warm clouds 
($T\approx 7,000$~K) embedded
in a hot gas bubble created by supernova events, which likely occured
in the Lower
Centaurus Crux subgroup of the Scorpio-Centaurus Association 
during the last 10~Myr \citep{mai01}. 
The Local Bubble was last reheated and presumably shocked by a
supernova event 1--2 Myr ago \citep{lyu96,ber02}. 
If the Local Interstellar Cloud is representative \citep{red00},
typical column
densities for the warm clouds in the Local Bubble are $N({\rm H~I})
\sim10^{18}$~cm$^{-2}$ (see Table 2).  The origin of
these clouds is uncertain.  It is possible that they consist of material that
has cooled after being heated by thermal conduction or turbulent
mixing with hot gas.  In this case some fraction of deuterium in
grains might have been returned to the gas by thermal sputtering
while the gas was sufficiently hot for this to be effective.
Without knowledge of the thermal history of individual Local
Bubble gas clouds, it is not possible to predict what values of
(D/H)$_{\rm gas}$ are expected in this interpretation, but the
observed value of $\sim$15~ppm appears to be compatible with this
picture.  Alternatively, pre-existing neutral clouds might have
been shocked by low-velocity shocks driven into the clouds by a
sudden increase in ambient pressure.  \citet{fri99} argue that the
abundances of Si, Mg, and Fe in the Local Interstellar Cloud are 
consistent with grain processing through a shock with $\sim 100$ km~s$^{-1}$. 
In this interpretation,
grain-grain collisions would be expected to result in destruction
of only a few percent of the refractory grain material like iron
\citep{fri03}. Since the condensation temperature \citep{sav96}
of Fe (1300~K) is much
larger than for C ($\sim 50$~K) to which the D is bound, we anticipate that 
grain-grain collisions, radiation, or other heating processes would strip
D from grains before Fe.
Since, as we will later argue, the total (gas and dust) value of (D/H) in the 
local Galactic disk is about 45\% larger than (D/H)$_{\rm LBgas}$, 
we surmize that 
the shocks in the Local Bubble have only partially evaporated the 
deuterium-bearing grains that could have been formed in molecular clouds 
before the onset of star formation in the Association.
As we shall see in \S6.1, the modest depletion of deuterium in the Local 
Bubble corresponds to modest depletions of the refractory metals iron and  
silicon and, as described in \S2, also oxygen. 

     Radio observations in the H~I 21~cm line provide the best information on
the properties and number density of individual diffuse H~I clouds in the ISM.
Typical diffuse H~I clouds have column densities in the range 
$\log N({\rm H~I}) =$ 18.8--19.8 with a median column density 
$\log N({\rm H~I}) = 19.3$
\citep{loc86,loc02}. \citet{cox87} cite a mean temperature for diffuse H~I
clouds of $\sim$80~K. They also say that, on average, the number of such clouds
per kpc in the disk with column densities in excess of $N({\rm H~I})$ is 
$5.7[N({\rm H~I})/10^{20}\: {\rm cm}^{-2}]^{-0.8}$. 
Using this formula, the number of clouds per kpc with
$\log N({\rm H~I}) > 19.3$ is about 20, with a typical cloud 
separation of about 50
pc; however, the distribution should be very patchy. 

     Lines of sight extending beyond the Local Bubble should pass though either
recently shocked gas, not recently shocked gas, 
or a mixture of such regions. According to our
time-dependent depletion model, the highest measured values of 
(D/H)$_{\rm gas}$ should indicate recently shocked regions where 
(D/H)$_{\rm gas} \leq$ (D/H)$_{\rm total}$, and the
lines of sight with low values of (D/H)$_{\rm gas}$ should indicate regions 
that
have not been shocked for a long time and thus have severe deuterium depletion.
Lines of sight extending beyond the Local Bubble 
with the highest measured values of  
(D/H)$_{\rm gas-LB}$, that is values above 21 ppm, include the lines of sight 
toward $\gamma^2$ Vel, Lan~23, WD~1034+001, Feige 110, and LSE~44. 

     Given the patchy distribution of diffuse H~I clouds, lines of sight
extending beyond the Local Bubble can include none, one, or many of these
clouds. If there is, at most, one diffuse H~I cloud in a line of
sight, then (D/H)$_{\rm gas}$ should reflect the conditions of warm 
interstellar gas near the Sun 
that is partially ionized and likely has not had sufficient time to deplete
an appreciable fraction of deuterium onto dust after the last shock. However,
when a line of sight passes through several or many diffuse H~I clouds, 
the composite value of (D/H)$_{\rm gas}$ should be dominated by the abundances
in these clouds where gas-phase deuterium
can be highly depleted onto the cold grains even when the gas is not cold.
Thus lines of sight in the intermediate regime ($\log
N({\rm H~I}) =$ 19.2--20.7) could contain little or no unshocked H~I gas or be
dominated by such material, leading to a wide range of 
(D/H)$_{\rm gas}$ values. 

{\it HST} and {\it Copernicus} spectra of $\gamma^2$ Vel 
and diffuse H$\alpha$ emission near the star
show highly ionized gas (including O~VI)
along its line of sight from which \citet{fit94} and \citet{rey97} infer an 
H~II region of path
length 100--150~pc produced by a wind-blown superbubble and photoionizing 
radiation from $\gamma^2$~Vel and other massive stars.
These conditions of strong ultraviolet radiation and recent 
supernova-produced shocks could remove deuterium
from the mantles of dust grains, leading to the observed high value of 
(D/H)$_{\rm gas}$. The ionization along the other lines of sight with high 
(D/H)$_{\rm gas}$ is not yet characterized.


For lines of sight with larger column densities ($\log N({\rm H~I}) > 20.7$),
we expect that most of
the column density will be contributed by cool H~I gas, and the
overall (D/H)$_{\rm gas}$ value should be close to the low
(D/H)$_{\rm gas}$ values prevalent in the cool H~I phase. 
Measurements of (D/H)$_{\rm gas}$ along as many long lines of sight as possible
with {\it FUSE} and other instruments are needed to test this hypothesis. 
For a discussion of possible line saturation effects see \S~6.3.2.

\section{TESTING THE DEUTERIUM DEPLETION MODEL}

     We now pose two tests to determine whether or not deuterium 
depletion is a valid model for explaining the wide range of
(D/H)$_{\rm gas}$ values for lines of sight extending beyond the Local Bubble.
While neither test is conclusive, the statistical trends identified by
these tests taken 
together make for a strong case. One problem in this analysis is
that the lines of sight to the very bright stars
analyzed with data from {\it Copernicus} have
larger uncertainties in (D/H)$_{\rm gas}$ than the lines of sight 
to a different group of stars studied
using {\it IMAPS} or {\it FUSE} spectra. {\it IMAPS} reobserved the
{\it Copernicus} stars $\gamma^2$~Vel, $\zeta$~Pup, and $\delta$~Ori, 
but {\it FUSE} was unable to reobserve 
any of the {\it Copernicus} stars to obtain more 
accurate D/H measurements. A rough estimate of the improvement in the 
measurements obtained with {\it IMAPS} and {\it FUSE} comes from a 
comparison of the (D/H)$_{\rm gas}$ values for $\gamma^2$ Vel: the 
{\it Copernicus} result \citep{yor76} was 
$17.8^{+10.4}_{-6.6}$ ppm, whereas the {\it IMAPS} result \citep{son00}, 
which we use in 
our analysis, was $21.8\pm 2.1$ ppm. Although the two results are 
consistent within the {\it Copernicus} error bars, the {\it IMAPS} value 
is 2.1$\sigma$ ({\it IMAPS} error bars) above the value obtained 
from {\it Copernicus} spectra. \citet{vid77} used {\em Copernicus} data
to infer two velocity components in the line of sight to $\zeta$~Pup:
in component 1 (D/H)$_{\rm gas} = 22$~ppm with no error estimate, and in
component 2 (D/H)$_{\rm gas} = 23^{+7}_{-3}$~ppm. The high spectral
resolution and S/N of the {\it IMAPS} data allowed \citet{son00} to 
resolve or nearly resolve 
the complex velocity structure for this line of sight and to infer
(D/H)$_{\rm gas} = 14.0\pm 2.3$~ppm, which is well below the 
{\em Copernicus} result. 
On the other hand, for the $\delta$~Ori line of sight there is good agreement
between the {\it Copernicus} result of $7.5\pm 2.3$~ppm \citep{lau79} and the
{\it IMAPS} result of $7.4^{+1.2}_{-0.9}$~ppm \citep{jen99}.
These comparisons suggest that we should be 
careful when 
including {\it Copernicus} data together with the more accurate recent
data. In particular, the {\it Copernicus} data for the lines of sight to
$\alpha$~Vir and $\mu$~Col have uncertainties
in (D/H)$_{\rm gas}$ that are near 100\%. We have chosen to include all of the 
{\it Copernicus} data, which will clutter the subsequent figures somewhat, but
in computing correlations we weight each of the data points by the inverse 
of the $1~\sigma$ range in (D/H)$_{\rm gas}$. In this way, the recent data 
with small measurement uncertainties will dominate the correlation fits.

     In Figures 3--8 we compare (D/H)$_{\rm gas}$ vs. other observables
along the same lines of sight. Except for Figure~5, we do not use 
(D/H)$_{\rm gas-LB}$ because we have no reliable way of removing the Local
Bubble foreground from the other quantity that (D/H)$_{\rm gas}$ is being
compared with for most lines of sight. For all of the correlation tests,
we compute weighted least-squares linear
fits, log[$N$(D~I)/$N$(H~I)] = a + bD(Fe) or a + bD(Si), 
to the observed data points by allowing all of the data quantities,
$N$(H~I), $N$(D~I), and either $N$(Fe~II) or $N$(Si~II), to vary randomly 
within their 1$\sigma$ error range. 
We compute a and b parameters for 1000 such realizations 
and then fit Gaussians to the distributions of the a and b parameters. Table~5
lists the mean values of the a and b parameters and their 1$\sigma$ 
uncertainties,
which are the half-widths of the corresponding Gaussian distributions.  
The relatively small 1$\sigma$ uncertainties in the b slope parameters for 
these fits demonstrates that correlations are real despite the 
correlation of the errors in the x-axis and y-axis quantities 
(see \S6.3.6 for a different approach). 
For each plot we also apply the Spearman rank correlation test\footnote{We use 
the IDL library routine R\_CORRELATE
to compute the Spearman ($\rho$) rank correlation test. The description 
of this test is from the IDL Reference Guide.},
which provides an unbiased means for determining whether the null 
hypothesis of no correlation can be rejected for the unweighted data points.
Column~9 in Table~5 gives the Spearman rank correlation coefficient, and 
column~10 gives the two-sided significance of its deviation from zero. Small 
values of this significance parameter indicate significant correlation. 
The quantity in parenthesis is the 
number of standard deviations by which the significance parameter deviates 
from its null-hypothesis value.
These quantities are listed in Table~5 for the various tests 
described below.

\subsection{\it First Test: Correlation with Refractory Metal Depletions}

     \citet{dra04, dra06} proposed that a good test of whether 
deuterium depletion
onto dust is responsible for the observed variations in (D/H)$_{\rm gas}$ is
to see whether there is a positive correlation of depletion of metals like
Fe, Si, and Ti with (D/H)$_{\rm gas}$. Therefore, our
first test of the validity of the deuterium depletion model is to
determine whether for the same lines of sight, the observed (D/H)$_{\rm gas}$ 
ratios correlate with the depletion of refractory metals, that is, elements 
that easily condense onto interstellar grains because of their high chemical 
bonding energies.
In the low density ISM, the primary source of grain erosion is due to 
high-energy-sputtering processes that occur in
supernova-generated shock waves (Jones 2000). Since iron in grains is very
resistant to sputtering, small values of the iron depletion would indicate 
significant processing of the grains that, according to the model, 
would predict that
most or all of the deuterium has been removed from the grains. Gas-phase iron 
abundances can be measured accurately in the ISM, as {\it STIS} and {\it FUSE}
spectra both provide accurate measurements of the dominant ionization stage, 
Fe$^{+}$, for both cold and warm gas. The depletion of iron, 
D(Fe) = $\log_{10}[({\rm Fe/H})_{\rm gas}/({\rm Fe/H})_\odot]$, 
varies from typically $-1.2$ in warm diffuse clouds to typically $-2.2$ in cool
diffuse clouds (Savage \& Sembach 1996; Jenkins 2004). 
Very small iron depletions would
indicate that the dust grains have been vaporized, depositing iron and
deuterium from the dust back into the gas phase. 
Thus, very large negative values of
D(Fe) should correlate with low (D/H)$_{\rm gas}$, if the deuterium depletion
model is valid. Conversely, small negative values of D(Fe) should
correlate with high (D/H)$_{\rm gas}$ measurements. Column (8) in Table~3 
lists the values of D(Fe) obtained using $N$(Fe~II) either from the reference 
given in column (11) of Table~2 or from
the data compilations of Jenkins, Savage, \& Spitzer (1986) and Redfield \&
Linsky (2002). In a few cases, we have revised slightly the values of
$N$(Fe~II) as a result of more accurately determined f-values \citep{how00}
for the optically thin lines that most strongly constrain the Fe~II column 
density. These revisions range from
+0.04 dex for BD+39$^{\circ}$3226 to --0.04 for $\theta$~Car.
We then compute D(Fe) using our tabulated value of $N$(H~I).
The undepleted value for iron is assumed to be the solar
abundance ratio, log (Fe/H)$_{\odot} = -4.55\pm 0.05$ \citep{asp05},
determined using a 3D hydrodynamical model of the solar atmosphere. 

     Most lines of sight inside the Local Bubble show iron depletions in the
range of $-0.7 > {\rm D(Fe)} > -1.3$, 
which are typical for warm diffuse clouds.
Two lines of sight (36 Oph and G191-B2B) show smaller iron depletions and have
values of (D/H)$_{\rm gas}$ that are typical for the Local Bubble 
($14.1\pm 5.8$~ppm and $16.6\pm 4.1$~ppm, respectively). 
There is no measurement of D(Fe) for
the line of sight to $\beta$~Cet, but magnesium is undepleted for this line
of sight (Piskunov et al. 1997). 
Also, silicon, another refractory metal, is not depleted on this
line of sight.  From \citet{jen04}, one sees that D(Fe) $\approx$
D(Si) -- 0.8, so we might estimate D(Fe) $\approx -0.8$ for the
line of sight to $\beta$~Cet, which also has the highest value of
(D/H)$_{\rm gas} = 21.9^{+5.2}_{-6.8}$~ppm in the Local Bubble.  This
connection of high (D/H)$_{\rm gas}$ with minimal metal depletion is
as predicted by the deuterium-depletion model.

     In Figure~3, we plot (D/H)$_{\rm gas}$ vs. D(Fe) for all 38 lines of 
sight 
in Table~3 with measured D(Fe) values (excluding the upper and lower limits). 
Figure~3 shows a clear trend of decreasing (D/H)$_{\rm gas}$ with increasing 
iron depletion, especially for very large Fe depletions, D(Fe) $<-1.5$. 
The solid line in the plot is the least-squares-weighted linear 
fit to the data as described above.
We fit the data with an equation of the form 
(D/H)$_{\rm gas}$ = (a$\pm 1\sigma$) + (b$\pm 1\sigma$) D(Fe). 
For this fit,  the parameters are a$= 30.00\pm 1.60$ and
b$= 13.04\pm 1.15$. 
The Spearman rank correlation test\footnote{In \S6.3.5 we redo this analysis 
using only the data obtained with the {\em STIS}, {\em GHRS}, 
{\em IMAPS}, and {\em FUSE} instruments. The
effect is to remove the data points with the largest errors, leaving
data points that mostly have similar errors and thus similar weights.
The result for 26 data points is a similar linear fit that rejects
the null hypothesis of no correlation with $2.3\sigma$ significance.} of the  
unweighted data points shows a highly significant correlation parameter
and rejects the null hypothesis with 2.9$\sigma$ significance. 
These parameters are listed in Table~5.
While there are five
data points that lie above the linear fit, the statistical correlation 
of large Fe depletions
with small values of (D/H)$_{\rm gas}$ supports 
the deuterium-depletion hypothesis\footnote{There is no theoretical 
reason for assuming that the best-fit
relation should be linear, or that the D depletion and Fe
depletion should be tightly correlated with one another.}.
In \S~6.3 we discuss possible causes of scatter about the trend line,
which in particular make $\gamma^2$~Vel no longer appear 
anomalous. 

     Figure 4 is similar to Figure~3 but with the depletion of silicon D(Si)
for the x axis. The values of D(Si), listed in column (9) of Table~3, were
obtained using $N$(Si~II) from the references and the tabulated values of 
$N$(H~I). More significant revisions in $N$(Si~II) are needed than for Fe~II
as the f-values used for the weak Si~II 1020.7 and 1808.0~\AA\ lines have 
changed greatly since the {\em Copernicus} and {\em IUE} eras. 
We use the \citet{mor03}
f-values for these lines, which are based on laboratory measurements. The 
changes in $N$(Si~II) range from +0.43 dex for BD+39$^{\circ}$3226 to --0.02 
dex for $\beta$~CMa with most changes significantly positive.
We compute the silicon depletion assuming the solar abundance of silicon as 
log (Si/H)$_{\odot} = -4.49\pm 0.04$ \citep{asp05}. We find that D(Si), 
like D(Fe), shows a clear trend of increasing depletion with lower deuterium 
abundance. Our weighted least-squares fit to the 20 data 
points (excluding upper limits) is of the form
(D/H)$_{\rm gas}$ = (a$\pm 1\sigma$) + (b$\pm 1\sigma$) D(Si). 
The Spearman rank correlation test of the unweighted data points rejects the 
null hypothesis with $1.4\sigma$ significance. 
The parameters for this fit are given in Table~5.

     In their recent paper, \citet{pro05} find a similar correlation 
of (D/H)$_{\rm gas}$ with the depletion of titanium, which also has a high 
condensation temperature for the lines of sight to 
seven of the stars listed in Table~2 located beyond the Local Bubble.
Although Ti~II has a similar ionization energy to H~I and D~I, the 
velocity profiles of the Ti~II lines do not track very well the 
profiles of O~I 1335~\AA\ for the line of sight to HD~195965,
perhaps due to different depletion levels as suggested by \citet{pro05}.
Although the correlations of (D/H)$_{\rm gas}$ with D(Fe) and D(Si) are 
significant, the scatter that we have seen in Figure~3 and Figure~4
suggests that different physical processes are responsible 
for the depletions of D
compared to the metals and most likely among the different metals.
This could be the cause of the differences between the O~I and Ti~II line
profiles for HD~195965 and
would suggest significant scatter in plots of (D/H)$_{\rm gas}$ vs D(Ti).

     \citet{wak00}, \citet{jen86}, and others have demonstrated that the 
depletion of iron and other refractory 
elements becomes larger with increasing $N$(H~I), or increasing 
mean hydrogen number density, $\langle n_{\rm H}\rangle$,
along the line of sight. This relation
is generally explained in terms of the increasing contribution of quiescent 
cold clouds, 
with their large values of $N$(H~I) and greater metal depletions, to the
total $N$(H~I) when a line of sight passes through one or more such cold 
clouds. See also \citet{oli06} for a study of the correlation of 
(D/H)$_{\rm gas}$ with $\langle n_{\rm H}\rangle$ along the line of sight.
The correlations of (D/H)$_{\rm gas}$ with D(Fe) 
shown in Figure~3 and with D(Si) shown in Figure~4 are consistent 
with D, Fe, and Si being depleted onto grains in cold clouds.

\subsection{\it Second Test: H$_2$ Rotational Excitation}

The second empirical test of the deuterium depletion hypothesis is 
whether or not there is a 
correlation of (D/H)$_{\rm gas}$ with H$_2$ rotational temperature.
Table 6 summarizes the molecular hydrogen data for the fourteen lines of sight
with (D/H)$_{\rm gas}$ measurements. When both $N({\rm H_2})$ and $N$(HD) are
measured for a given line of sight, we include these molecules in the 
value of (D/H)$_{\rm gas}$. With some exceptions, for
column densities exceeding $\log N({\rm H})\approx 20.7$ 
(in our Galaxy), many H$_2$ spectral lines become 
optically thick, and the 
resulting shielding of the FUV radiation field decreases the photodissociation
rate leading to a high molecular hydrogen fraction, $f({\rm H}_2)$
\citep{spi75,sav77}. 
Under these conditions cloud densities can be sufficiently high that
the H$_2$ rotational excitation temperature $T_{01}$, derived
from the J=1 and J=0 levels of H$_2$, measures the gas kinetic temperature.
In their analysis of {\it FUSE} spectra of 129 sightlines to O and B stars in 
the Galactic plane, \citet{shu05} find that $T_{01} = 86\pm 20$~K. 
\citet{gil06} find that $T_{01} = 124\pm 8$~K characterizes lines of 
sight through the halo. For the Galactic plane data set, 
$\log [N({\rm H~I}) + 2N({\rm H}_2)] \geq 20.9$,
while our data set is mostly 
for smaller column densities. The three lines of sight in Table~6 
with the highest
column densities overlap the \citet{shu05} data set, and their values of 
$T_{01}$ are consistent with the \citet{shu05} range as expected.

   Eleven of the fourteen lines of sight listed in Table~6 have column 
densities  $\log [N({\rm H~I}) + 2N({\rm H}_2)]$ 
much smaller than  20.9 and, for the most 
part, values of $T_{01}$ much larger than 86~K. Since this is a new region
of parameter space to explore, we decided to test for a correlation
between $T_{01}$ and (D/H)$_{\rm gas}$.
Given that there is essentially no H$_2$ located inside the Local Bubble,
we plot (D/H)$_{\rm gas-LB}$ vs. T$_{01}$. The data shown in
Figure~5 do indeed show a correlation with one discrepant point, 
$\delta$~Ori, which is located behind a cloud that has been 
excited by stellar winds or explosive events \citep{jen00b}.
Our weighted least-squares fit to the 16
data points is of the form 
(D/H)$_{\rm gas-LB}$ = (a$\pm 1\sigma$) + (b$\pm 1\sigma$) 
log (T$_{01}$).
The Spearman rank correlation test of the unweighted data rejects the null 
hypothesis with $2.2\sigma$ significance, indicating a 
credible correlation. The fit parameters are listed in Table~5.
Although the correlation is only at a $2.2\sigma$ confidence level, we 
should ask why there should be any correlation at all. Since the lines of sight
are generally inhomogeneous, most of the molecular hydrogen along a given 
line of sight should be located in cool or cold clouds and thus not be 
cospatial with most of the atomic hydrogen and deuterium. Thus the properties
of H$_2$, such as $T_{01}$, could be independent of the depletions of D, Fe, 
and Si averaged over the line of sight. 
Nevertheless, we have found a correlation of $T_{01}$ with (D/H)$_{\rm gas}$ 
and therefore indirectly with the depletions of Fe and Si. 
Furthermore, for $N({\rm H}) < 20.7$,
$T_{01}$ should not be a good measure of the gas temperature but is rather 
determined by the balance of excitations and 
de-excitations of the lower levels of H$_2$ with collisions likely unimportant.
We note that in Table~6 all of the lines of sight with high values of $T_{01}$
have very small fractional abundances of H$_2$, $f$(H$_2$). For these lines 
of sight there is likely very little cold gas and thus less 
depletion of deuterium and metals. We do not know whether this is
a good explanation for the correlation of (D/H)$_{\rm gas}$ with
$T_{01}$ shown in Figure~5. Observations of more lines of sight with 
low values of $N$(H) but containing H$_2$ and further theoretical work are 
needed to verify the correlation and to explore other possible explanations.

\subsection{\it Possible Causes for Scatter and Discrepant Lines of Sight}

     Figure 3 shows that five lines of sight ($\gamma^2$~Vel, 
Lan~23, WD~1034+001, PG~0038+199, and TD1~32709) have high values of 
(D/H)$_{\rm gas}$ as compared 
to the general trend of the other lines of 
sight. In Figure~5, the line of sight to $\delta$~Ori has a low value of
(D/H)$_{\rm gas-LB}$ compared to the trend line. What could explain these 
discrepant data points and, for that matter, the scatter of data about the 
least-squares trend lines? We list here several considerations that may play 
important roles in explaining the scatter, but this list may not be complete:

\subsubsection{Inhomogeneous Lines of Sight}

     Until now, we have assumed that the 
properties of interstellar gas along a given line of sight have constant 
values inside the Local Bubble and then a different set of constant values
beyond the Local Bubble. This very simple approximation is invalid for most 
lines of sight that have been studied with sufficient spectral resolution to 
identify the different velocity components and their individual properties. If
the properties of interstellar gas in the different components are 
significantly 
different, for example in temperature, ionization (H~II vs. H~I regions), and 
density, then different observables may be formed preferentially in different 
locations and thus not correlate well. One example is the
correlation of (D/H)$_{\rm gas-LB}$ with the temperature of H$_2$, as H$_2$ 
could be 
present in only one of the components along the lines of sight. 
For lines of sight with low H$_2$ column densities, H$_2$ could be found in
relatively warm gas associated with H~I and neutral metals. For example,
high-resolution spectra of $\delta$~Ori obtained with the {\it IMAPS} 
instrument show H$_2$ line profiles with similar
velocity structure as N~I, O~I, and D~I \citep{jen99,jen00b}.
This could explain why $T_{01}$ is large for the the $\delta$~Ori 
line of sight.

     One way of estimating the depletion of an element in a particular 
velocity component without information on $N$(H~I) for that component is to 
reference the gas-phase column density of the element to another element
like sulphur, which is not significantly depleted in H~I regions. 
\citet{how99} have used 
this approach to derive D(Fe) and D(Si) for Component 1 in the line of 
sight to $\mu$~Col using {\em GHRS} echelle spectra. This low-velocity 
component, also called Component A by \citet{shu77}, contains most of the 
neutral gas and H$_2$ along the line of sight. Tables~2, 3, and 6 include the
\citep{how99} data for Component 1.

\subsubsection{Line Saturation}

For complex lines of sight with large $N$(H~I), it is important to measure 
$N$(D~I) from the higher lines in the Lyman series to avoid line saturation
that could lead to underestimates in $N$(D~I) and therefore in 
(D/H)$_{\rm gas}$.
The 20 km s$^{-1}$ spectral resolution of {\it FUSE} is inadequate
to resolve the D lines or to separate velocity components in the line. 
In particular, narrow components from cold gas could be saturated but not
recognizable at the resolution of {\it FUSE}. 
We have argued in \S~5 that the low values of (D/H)$_{\rm gas}$ for lines of 
sight beyond the Local Bubble could be due to the 
dominant role of cool diffuse clouds in which deuterium is largely depleted.
Such clouds would have narrow absorption features observable in metal lines.
Indeed, \citet{red02} list a number of lines of sight with Ca~II absorption
features with $b<1.5$~\kms. The chances of this occuring
grow with increasing $N$(H~I), but measurements of $N$(D~I) in 
high members of the Lyman series, which require high S/N data, 
decrease the probability that line saturation will occur.
One can test for D line saturation by looking for an increase in $N$(D~I)
toward higher lines in the Lyman series or from a curve of growth analysis
(e.g., Friedman et al. 2002). Also, high resolution spectra
of interstellar metal lines can identify velocity features that could have
saturated D absorption. The
blending of adjacent Lyman lines with each other and with 
the many H$_2$ lines will eventually 
set a limit to the largest value of  $N$(H~I) for which one can measure 
unsaturated D Lyman lines,
but this should not occur until $\log N({\rm H~I})$ is considerably larger 
than 21.0 \citep{jen96}.

\subsubsection{Ionization Corrections}

      \citet{jen86} and \citet{jen04} have argued that anomalously small
metal depletions compared to typical lines of sight with the same $N$(H~I) 
could be due to H II regions where hydrogen is partially ionized but 
metals with second ionization potentials greater than 13.58~eV, such as
iron, silicon, and phosphorus \citep{leh04}, remain mostly 
singly ionized. 
Positive values of D(Si) are consistent with Si$^+$ being the
dominant ionization state and H being partially ionized. A good example of 
this is the line of sight to $\beta$~CMa. If we assume that essentially
all of the hydrogen is H~I and all of the silicon and iron are singly ionized
for this line of sight,
then D(Si)= $+0.41\pm 0.16$, and D(Fe) = $-0.30\pm 0.16$ \citep{gry85, dup98}.
However, along this line of sight hydrogen is mostly ionized \citep{gry85}. 
In their analysis of this line of sight,
\citet{dup98} identify four velocity components with most of the H~I column 
density in Component C. We therefore list in Table~3 the depletions for 
Component C that include all ionization stages for hydrogen, silicon, and
iron, D(Si) = $+0.04^{+0.20}_{-0.14}$ and D(Fe) 
$=-0.61^{+0.20}_{-0.14}$ \citep{dup98}.
The line of sight to $\beta$~Cet also has a positive value for the
silicon depletion, D(Si) $= +0.30\pm 0.41$.
\citet{red04b} showed that one of 
the two velocity components toward $\beta$~Cet has a very high temperature, 
$T=12,400\pm 2,800$~K, likely indicating partial ionization of hydrogen. 
Thus ionization corrections of presently unknown size 
should be applied to
D(Si) for this line of sight, which would move its data point to the 
right in Figure~4.  The high 
temperature of the gas toward $\beta$~Cet is likely due to a recent shock, 
which is consistent with the high value of 
(D/H)$_{\rm gas} = 21.9^{+5.2}_{-6.8}$ ppm for this line of sight.

     The $\gamma^2$~Vel line of sight has seven velocity components -- three 
are H~II regions, likely produced by ionization from the Wolf-Rayet star 
in the binary system, 
and four are H~I regions \citep{fit94}. The H~I regions 
contribute 89\% of $N$(Fe~II), but the fractional ionization of hydrogen 
and $N$(H~I) in each component is not known. The iron depletion listed 
in Table~3 was therefore computed from the line-of-sight
integrated values of $N$(H~I) and $N$(Fe~II), rather than from the column 
densities in the H~I components. Inclusion of the unknown 
$N$(H~II) would shift D(Fe) to the right in Figure~3. Since sulfur is not 
usually depleted, we estimate that 72\% of the total hydrogen column density 
is located in the H~II regions by summing the column densities of S~I, S~II, 
and S~III in each velocity component. 
Although the inclusion of this amount of ionized hydrogen would shift
D(Fe) for $\gamma^2$~Vel by 0.55 dex to the right in Figure~3, an unknown
but considerable amount of Fe could be doubly ionized, shifting D(Fe) by an
unknown amount to the left. Realistic calculations of the ionization of 
H and Fe are needed to address this problem.

     Ionization corrections can be important for lines of sight with small 
values of $N$(H~I) because ionizing radiation can then penetrate the gas 
and preferentially ionize H compared
to Fe$^+$ and Si$^+$. For example, \citet{fri03} find in their ionization
equilibrium models for the nearby ISM that hydrogen is about 30\% ionized
while 96.5\% of Fe is Fe~II and 99.6\% of Si is Si~II.
To avoid lines of sight where such
ionization corrections are likely important, we reconsider the correlation of
(D/H)$_{\rm gas}$ with D(Fe) and D(Si) but now remove the data for 
lines of sight with  $\log N({\rm H~I})<19.0$. 
This removes all but two of the Local Bubble lines
of sight. In Figure~6, we plot (D/H)$_{\rm gas}$ vs. D(Fe) 
for the 24 data
points that meet this criterion. The slope is now steeper than for the full
data set (see Table~5). The Spearman rank
correlation test rejects the null hypothesis of no correlation with
$2.6\sigma$ significance. 

     Figure~7 is similar to Figure~4, except that we remove the data points 
with $\log N({\rm H~I})<19.0$ 
and plot (D/H)$_{\rm gas}$ vs. D(Si) 
for the 11 data points that meet this criterion. 
The slope is nearly the same as for the full
data set (see Table~5). The Spearman rank
correlation test rejects the null hypothesis of no correlation with
$2.1\sigma$ significance. The data are in excellent agreement with the
linear fit, except for the one uncertain data point, $\mu$~Col.

     The removal of lines of sight that are most likely affected by ionization
corrections leads to steeper correlations of (D/H)$_{\rm gas}$ with 
D(Fe) and 
D(Si), fewer discrepant points, and tighter correlations. The remaining scatter
could be due to measurement errors (especially for some of the 
{\em Copernicus} data), inhomogeneous lines
of sight, and different grain compositions. The ionization of H and D should be
the same as they share the same bound-free continuum.

\subsubsection{Different Grain Compositions}

      One should not expect a simple one-to-one relation between 
(D/H)$_{\rm gas}$ and D(Fe) or D(Si) 
along a given line of sight, because deuterium and different 
metals may be depleted onto 
different types of grains or be located preferentially in different layers of
the same grain. 
Many authors (e.g., Spitzer \& Fitzpatrick 1993) have argued that grains
typically have hardy cores but more easily destroyed mantles.
For example, iron may be concentrated in the grain cores and 
some deuterium concentrated in PADs either free-flying or 
incorporated into larger grains. 
\citet{jon94} estimated the
destruction timescale for grains by supernova-driven shocks to be
2--3 $\times 10^8$ yr but the timescale for the 
erosion of grain mantles by weaker shocks is much shorter. 
We note that the five lines of sight that
lie above the trend line in Figure~3 ($\gamma^2$~Vel, Lan~23, WD~1034+001,
PG~0038+199, T1~32709)
also have the highest values of (D/H)$_{\rm gas}$ (see Figure~1). (Feige~110, 
the fourth line of sight with a high value of (D/H)$_{\rm gas}$, does not have 
D(Fe) or D(Si) measurements.) Clearly, there is no simple 
one-to-one relationship 
between (D/H)$_{\rm gas}$ and metal depletion, although there are the
general trends indicated by the least-squares trend lines. In Figure~6, where 
we have considered only lines of sight with $\log N({\rm H~I}) > 19.0$ to 
minimize
ionization corrections, all of the five high lines of sight are still present.

In Figure~3 the dashed line is drawn parallel to the least-squares fit to 
all of the
lines of sight (solid line) but arbitrarily displaced upwards by 8.5~ppm.
The dashed line is a good fit to the five high data points. The high values of
(D/H)$_{\rm gas}$ for these five lines of sight could be explained by either
a smaller than usual percentage of the grains being composed of carbon and 
thus few sites available for deuterium to deplete on
or the evaporation of deuterium that was in grain mantles by weak shocks.
If the slope of the dashed line is real and not an artifact of the 
few data points, then there is support for a smaller than usual percentage of 
carbon grains in these lines of sight. Since the condensation temperature
for carbon is much smaller than that for iron (typically in the form of oxides)
\citep{sof94,sav96}, a low percentage of carbon grains compared to 
iron grains is likely for some lines of sight. 
 
\subsubsection{Difficulty of Measuring N(H~I)}

The most difficult column density to measure accurately 
is usually hydrogen because the
Lyman lines are very optically thick, they must be measured against an 
uncertain stellar background (Lyman absorption or emission lines), 
and they are very 
broad, which requires interpolation over a large wavelength range. 
The hydrogen column densities cited in Table~2 are based on careful analyses
that consider these and other effects. The most accurate values of $N$(H~I) 
are generally obtained from analyses of {\em STIS}, {\em GHRS}, and {\em FUSE}
spectra.  We note that $N$(H~I) for the Feige~110 line of sight was obtained 
from the analysis of one {\em IUE} spectrum. 
Since the Lyman-$\alpha$ line lies near the end of
the {\em IUE} spectrum where the sensitivity is low and the 
echelle spectral orders are close 
together, it is difficult measure the background and continuum. While there is
no evidence that the value of $N$(H~I) listed in Table~2 is more uncertain 
than the cited error bars, the Feige~110 line of sight should be 
reobserved with a better spectrograph 
when feasible. The Lan~23 line of sight should also be reobserved as $N$(H~I) 
was obtained from a noisy low dispersion {\em EUVE} spectrum rather than 
from Lyman line spectra. To test whether the less reliable values of $N$(H~I) 
are biasing our conclusions, we plot in Figure~8 only the data points for 
which $N$(H~I) was obtained from {\em STIS}, {\em GHRS}, or {\em FUSE}.
The fit to the data and the parameters characterizing the fit listed in 
Table~5 are similar to the results obtained when we included all of the data
in Figure~3.

\subsubsection{Removing a Possible Correlation from the Plots}

     In Figures~3-4 and 6-8, the measured $N$(H~I) values enter the quantities 
plotted in both the x and y-axes, raising the possibility of 
inaccurate or false correlations. To eliminate this possibility, 
we have plotted in Figure~9  
the $\log [N({\rm D~I})/N({\rm Fe~II})]$ ratio vs. $\log N({\rm H~I})$ 
for all lines of sight. 
We include in the figure the weighted least-squares fit to all of the data
points (solid line) and the fit to only the lines of sight which extend 
beyond the Local Bubble (dashed line). 
The essentially zero slope in the fit to the data beyond the Local Bubble 
indicates that the positive slope of the fit to the entire data set is 
determined by the low values of $\log [N({\rm D~I})/N({\rm Fe~II})]$ 
for lines of sight inside the Local Bubble. 
The zero slope beyond the Local Bubble indicates that on a statistical
basis the deuterium and iron depletions are proportional.
The scatter of the data about the two regression lines
suggests different grain compositions along the various lines of sight 
or different processes that deplete deuterium and iron from
the gas phase but that depletion is important for both species. 
While the scatter indicates that the depletions of D and Fe are not 
proportional on individual lines of sight, the zero slope fit to the data 
beyond the Local Bubble indicates that on average the rates of depletion 
of D and Fe appear to be proportional.

     We noted in \S6.3.3 that in regions of low shielding 
($\log N({\rm H~I}) <19.0$)
from Lyman continuum radiation primarily from hot stars,  
that hydrogen can be partially 
ionized, whereas metals with second ionization potentials $> 13.58$~eV 
will be mostly singly ionized. 
The second ionization potential of iron is 16.18~eV, 
whereas the ionization potential of neutral deuterium is the same as that of 
hydrogen. Thus partial ionization of deuterium could explain the decrease
in $N$(D~I)/$N$(Fe~II) at $\log N({\rm H~I}) < 19.0$. 
It may also explain the very low 
ratios for the short lines of sight to 36~Oph and G191-B2B. \citet{red00}
found that hydrogen is half ionized in the Local Interstellar 
Cloud, which has a maximum column density $\log N({\rm H~I}) = 18.3$. 
This result
could simply explain the factor of two decrease in $N$(D~I)/$N$(Fe~II) between 
lines of sight with large and small values of $N$(H~I) in Figure~9.

\section{WHAT IS THE TOTAL DEUTERIUM ABUNDANCE IN THE LOCAL GALACTIC DISK?}

     We propose that the most plausible explanation for the wide range in
(D/H)$_{\rm gas}$ measurements within about 1 kpc of the Sun is that 
different amounts of deuterium
depletion occur in different lines of sight. We have come to this conclusion
because: (i) theoretical arguments suggest that deuterium atoms can replace
enough hydrogen atoms in carbonaceous grain materials to explain low values of
(D/H)$_{\rm gas-LB}$ in the undisturbed ISM,
(ii) high (D/H)$_{\rm dust}$ ratios have been measured in 
interplanetary dust
particles that likely came from the ISM, providing an important proof of
concept for the deuterium depletion hypothesis, and (iii) the correlation of
large iron and silicon depletions in the gas phase, D(Fe) and D(Si), 
with low values of (D/H)$_{\rm gas}$
strongly support the hypothesis that dust grains are an important reservoir for
deuterium and that deuterium returns to the gas phase when the grains are
disturbed.

     The deuterium-depletion model predicts that the most likely value for
the ratio of deuterium in all forms (gas plus dust) to hydrogen, 
(D/H)$_{\rm total}$,
for the local region of the Galactic disk, (D/H)$_{\rm LDtot}$, should be 
{\em equal to or slightly} above the highest measured 
(D/H)$_{\rm gas}$ ratios. If some 
deuterium remains on grains for these lines of sight, then the total amount 
of deuterium will slightly exceed the gas-phase value. 
In Figure~1 there are five lines of sight ($\gamma^2$~Vel, Lan~23, 
WD~1034+001, Feige~110, and LSE~44) extending beyond the Local Bubble that 
have the highest values of (D/H)$_{\rm gas} \approx 22$~ppm. These are the 
best candidates for inferring (D/H)$_{\rm LDtot}$ in the local Galactic disk.
The weighted mean and standard deviation of the mean for these 
five data points 
is $\langle ({\rm D/H})_{\rm gas}\rangle = 21.7\pm 1.7$~ppm. 
However, the Local Bubble
foreground has a known mean value of (D/H)$_{\rm gas}$ and extent,
$\log N({\rm H~I})\approx 19.2$, indicating that  deuterium is depleted
inside the Local Bubble. We therefore subtract the Local Bubble 
foreground from the (D/H)$_{\rm gas}$ values to obtain the values of
(D/H)$_{\rm gas-LB}$ listed in Table~3. The weighted mean values for these
five lines of sight is 
$\langle ({\rm D/H})_{\rm gas-LB}\rangle = 23.7\pm 2.4$~ppm. 
The line of sight to $\gamma^2$~Vel 
lies closest to the Local Bubble ($\log N({\rm H~I})=19.710\pm 0.026$)  
and is thus most subject to the 
systematic errors (especially the uncertain extent of the Local Bubble 
in $N$(H~I))
associated with subtracting the Local Bubble contribution. However, this 
line of sight has the highest weight (58\%) in computing this mean. 
To minimize this potential source of systematic error, we have recomputed the
mean without weighting by the inverse errors, obtaining
$\langle ({\rm D/H})_{\rm gas-LB}\rangle = 23.1\pm 2.4$~ppm. 
Given that we have selected
the data points to be used in computing the mean from a larger sample, 
the errors are not Gaussian,
but we believe that our method gives the most representative value 
(D/H)$_{\rm gas-LB}$ available at this time. Since even for these five lines
of sight some deuterium could be in the grains, 
(D/H)$_{\rm LDtot} \geq \langle({\rm D/H})_{\rm gas-LB}\rangle = 
23.1\pm 2.4$~ppm.
 
Even the higher estimate of the primordial D/H ratio (see Section 1), 
(D/H)$_{\rm prim} = 27.5^{+2.4}_{-1.9}$~ppm,
implies a small deuterium astration (or depletion)
factor for the local region of the local Galactic disk, 
$f_d \equiv$(D/H)$_{\rm prim}$/(D/H)$_{\rm LDtot} \leq 
(27.5^{+2.4}_{-1.9})/(23.1\pm 2.4) \leq 1.19^{+0.16}_{-0.15}$. 
For the lower estimate of (D/H)$_{\rm prim} = 26.0^{+1.9}_{-1.7}$~ppm,
$f_d \leq (26.0^{+1.9}_{-1.7})/(23.1\pm 2.4) \leq 1.12\pm 0.14$
These small values of $f_d$ provide an
important constraint on models of Galactic chemical evolution,
which must also explain the evolution of many chemical species. 
The models of \citet{chi02} (cf. Romano et al. 2003), for example, predict
$f_d \simeq 1.5$ for a wide range of assumed (D/H)$_{prim}$ and 
different rates of extra mixing in stellar interiors. These models include
two epochs of infall of D-rich and metal-poor gas\footnote{The infalling gas
has a metal abundance 20\% solar \citep{rom03}, which is consistent with
the results for the high velocity cloud Complex C \citep{col03}.} 
from the halo or IGM, and 
the models predict sensible present epoch values for $^3$He/H, $^4$He/H, 
and $^7$Li/H. 
By altering the assumptions regarding the rates of infall and star formation,
more recent work by \citet{rom06}
is able to reproduce current data on
stellar abundance patterns in the solar neighborhood ($R=8$~kpc) 
with astration factors $1.3 < f_d < 1.8$; the lower end of this range
overlaps our determinations $f_d < 1.19^{+0.16}_{-0.15}$ or 
$f_d < 1.12\pm0.14$.

     There are also estimates of the D/H ratio in the protosolar cloud,
(D/H)$_{\rm psc}$, when the Galaxy was about two-thirds of its present age.  
\citet{gei98} have inferred 
(D/H)$_{\rm psc} = 21\pm 5$~ppm based on measurements of
$^3$He/$^4$He in the solar wind, measurements of the same quantity in
Jupiter's atmsphere, and the assumption that $^3$He in the solar atmosphere 
has not undergone nuclear reactions. 
Analysis of {\it Infrared Space Observatory (ISO)}
spectra of H$_2$, HD, CH$_4$, and CH$_3$D in Jupiter's atmosphere 
led \citet{lel01} to infer (D/H)$_{\rm psc} = 21\pm 4$~ppm, although there are
many assumptions that go into this analysis. One cannot directly compare
(D/H)$_{\rm psc}$ with present day estimates of (D/H)$_{\rm LDtot}$, 
because the Sun 
has moved a considerable distance from its place of birth about 4.6 Gyr ago.
However, the similar values of (D/H)$_{\rm psc}$ and 
(D/H)$_{\rm LDtot} \geq 23.1\pm 2.4$~ppm suggest that 
most of the decrease in the total deuterium abundance in the ISM 
may have occurred as
a result of nuclear reactions in stars and subsequent transfer of 
deuterium-depleted gas into the ISM during the first two-thirds of the 
age of the Galaxy.

\section{COULD INFALL EXPLAIN THE HIGH (D/H)$_{\rm gas}$ RATIOS?}

It is fair to ask whether depletion is the only operative mechanism for 
changing the abundance of deuterium. \citet{oli05} have used the 
constancy of the (O/H)$_{\rm gas}$ ratios over a range of environments in the 
local part of the Galaxy to argue that significant changes in the D/H 
ratios because of infall of lower metallicity gas
are unlikely.  While it is tempting to suggest 
significant and highly localized infall for the high D/H sightlines, 
such a large infall would dramatically lower the O I abundances 
along these sightlines. A large increase in the value of (D/H)$_{\rm gas}$ 
from 15 ppm to 22 ppm would require a highly localized infall of gas with 
(D/H)$_{\rm total}=24$ ppm within a mixing time of about 350 Myr
\citep{dea02}.  The fraction due to infall would equal more 
than three quarters of the mass of the gas along a given sightline. 
Such a large infall of D-rich, but metal-poor, gas would also lower the 
measured (O/H)$_{\rm gas}$ ratios by more than a factor of three; 
such low values are not observed. 
The constancy of (O/H)$_{\rm gas}$ compared to variations in 
(D/H)$_{\rm gas}$ thus
argues against localized infall \citep{oli05}. Although 
the infall mechanism is unlikely to explain the wide range of observed
(D/H)$_{\rm gas}$ values, we do not rule out the possibility 
of another, as yet unknown, mechanism for increasing the D abundance.

     Finally, we include a note of caution. \citet{heb05} have pointed out 
that the (D/O)$_{\rm gas}$ data show variations similar to those observed 
for (D/H)$_{\rm gas}$, except that there were no high values at that time 
corresponding to those measured for (D/H)$_{\rm gas}$.  The 
relative uniformity of O~I/H~I allows O~I to be used as a proxy for H~I, 
with the advantage that for a given line of sight, $N$(O~I) is much closer 
to $N$(D~I) than $N$(H~I). Therefore,
the D~I and the O~I transitions in the {\it FUSE} range have 
similar opacities, lowering the chances of systematic 
errors due to large differences in optical depth \citep{heb03}.  A 
gas-phase (D/H)$_{\rm gas}$ ratio of $23.1\pm 2.4$ ppm and the 
(O/H)$_{\rm gas}$ ratio of $343\pm 15$~ppm measured by 
\citet{mey98} implies (D/O)$_{\rm gas} = 7.02\times 10^{-2}$,
nearly twice the mean measured value of $(3.84\pm 0.16)\times 10^{-2}$
\citep{heb03}. 
Although such high values had not previously been published,
\citet{oli06} report 3 new sightlines (WD~1034+001, BD+39~3226, and TD1~32709)
with high values of (D/O)$_{\rm gas}$, 
consistent with (D/O)$_{\rm gas}\approx 7.02\times10^{-2}$.

\section{CONCLUSIONS}

     The main goal of the {\it FUSE} mission has been to obtain accurate
measurements of (D/H)$_{\rm gas}$ for many sightlines 
in the Milky Way Galaxy and beyond in order to measure 
(D/H)$_{\rm prim}$ and to obtain constraints on Galactic chemical evolution. 
{\it FUSE} has now obtained the data needed for accurate measurements of
(D/H)$_{\rm gas}$ much further out in the Galaxy and with higher precision 
than previous instruments. In our summary of the published measurements of 
(D/H)$_{\rm gas}$ obtained from {\it FUSE}, {\it HST}, {\it Copernicus}, and 
{\it IMAPS}, we find, following \citet{woo04}, that the (D/H)$_{\rm gas}$
measurements appear to fall into three distinct groups depending on the 
neutral hydrogen column density $N$(H~I) to the target star. For lines of sight
with $\log N({\rm H~I}) < 19.2$, that is within the Local Bubble, 
the mean value of 
(D/H)$_{\rm gas}$ is $15.6\pm 0.4$ ppm, where the uncertainty is the standard
deviation of the mean. At large column densities, $\log N({\rm H~I}) > 20.7$, 
we find that the five lines of sight have much
lower values of (D/H)$_{\rm gas}$ with a mean value of $8.6\pm 0.8$~ppm. 
In the intermediate regime there is a factor of 4--5 range in the 
high precision (D/H)$_{gas}$ measurements with
values as low as $5.0^{+2.9}_{-3.4}$~ppm ($\theta$~Car) and as high as 
$21.8\pm 2.1$~ppm ($\gamma^2$~Vel). Since this large range has been 
measured from spectra obtained by several instruments, it is not an 
instrumental artifact, but rather an observational result that requires an 
explantion. 

     Both large changes in the local astration rates and infall of D-rich and 
metal-poor gas from the halo or IGM cannot explain the large range 
in (D/H)$_{\rm gas}$ without
predicting large variations in the local ISM metal abundances, in particular 
(O/H)$_{\rm gas}$, which are not observed. 
Instead, we describe a time-dependent deuterium-depletion model in which
much of the deuterium in the ISM 
can reside on grains under the appropriate conditions.
The theoretical basis for this model is the lower zero-point energy of 
C--D bonds compared to\\ 
C--H bonds. When
the ISM is undisturbed for a long time, grains are very cold, and
deuterium can replace hydrogen on carbonaceous grains and perhaps other types 
of grains. We estimate that D/H in the grains can  
approach $\sim$0.2, thereby reducing
(D/H)$_{\rm gas}$ to 6~ppm or lower. When a region of the ISM is shocked or 
the gas comes close to a hot star, the grains 
are partially or fully destroyed, sending deuterium atoms
back into the gas phase. In this model, the value of (D/H)$_{\rm gas}$ for a 
given line of sight depends on the environment and past history of the grains. 

Evidence in support of this model includes the measurement of high D/H ratios
in interplanetary carbonaceous dust grains, which are likely interstellar in 
origin. Since the conditions required from deuterium depletion are similar to
those for depletion of metals like Fe, Si, and Ti in the ISM, the model 
predicts a correlation of depletions of these metals with low values of
(D/H)$_{\rm gas}$. We present data that strongly support the correlation for 
Fe 
and Si, and \citet{pro05} present the data in support of correlation with Ti. 
The model also predicts that (D/H)$_{\rm gas}$ should be correlated with
the recent thermal and ionization history of the gas, whereas the variable
astration hypothesis would not be expected to show such a connection; 
we find that (D/H)$_{\rm gas}$ has a significant positive correlation with
the rotational temperature of the molecular hydrogen on the line-of-sight,
further supporting the hypothesis that (D/H)$_{\rm gas}$ is affected by
interstellar processes acting on relatively short time scales.
We note that in the most recent discussion of Galactic chemical 
evolution models, \citet{rom06} conclude that ``depletion of deuterium 
on to dust grains is the most likely physical mechanism proposed so far 
to explain the observed dispersion in the local data.''

If this model is realistic, then the best estimate of the total deuterium 
abundance, (D/H)$_{\rm total}$ would be the highest measured values of 
(D/H)$_{\rm gas}$
because these would be for lines of sight with minimum depletion 
of deuterium. We note that there are five lines of sight in the intermediate
$N$(H~I) regime with high values of (D/H)$_{\rm gas}$. 
The weighted mean value for these 
four lines of sight is $21.7\pm 1.7$~ppm (standard deviation of the mean). 
Since there could be some deuterium
depletion even in these lines of sight, we conclude that the best estimate of
the total D/H in the local disk region of the Galaxy, after subtracting
the Local Bubble foreground column densities,
is (D/H)$_{\rm LDtot} \geq 23.1\pm 2.4$~ppm\footnote{Very recently 
\citet{rog05} detected the hyperfine ground-state transition 
of atomic deuterium at 327~MHz (92~cm) in 
the Galactic anticenter direction. They find that 
(D/H)$_{\rm gas} = 23\pm 4$~ppm 
($1\sigma$) or $23^{+15}_{-13}$~ppm (3$\sigma$), and estimate that the column 
density-weighted mean distance of the emitting gas is about 2~kpc. Since the
beamwidth is $14^{\circ}$, the beam averages an unknown amount of cloud and
diffuse gas in and above the Galactic plane. While it is difficult to compare 
this new result with our line of sight measurements over shorter paths, the 
value of (D/H)$_{\rm gas}$ obtained from the hyperfine transitions of D and H 
is consistent with our proposed value of (D/H)$_{\rm LDtot}$.}.
This new estimate is $\geq 58$\% 
higher than the value of 15~ppm recently used in some Galactic 
chemical evolution models (e.g., Romano et al. 2003), which is 
(D/H)$_{\rm gas}$ for the Local Bubble rather than (D/H)$_{\rm LDtot}$.

     Two methods for determining (D/H)$_{\rm prim}$ based on data from very 
different times in the early universe are now in agreement. Measurements of
(D/H)$_{\rm gas}$ in five quasar absorption line systems have a mean value of 
$27.8^{+4.4}_{-3.8}$~ppm \citep{kir03}. Analysis of data from the {\it WMAP}
and other cosmic microwave experiments yield $\Omega_b h^2 = 0.0224$. 
Depending on the adopted nuclear reaction rates,  
(D/H)$_{\rm prim} = 27.5^{+2.4}_{-1.9}$~ppm \citep{cyb03} or 
$26.0^{+1.9}_{-1.7}$~ppm \citep{coc04}.
      
     Galactic chemical evolution models attempt to explain the decrease
in D/H from the protogalaxy, for which D/H is presumed to be the primordial 
value, to the present epoch where D/H = (D/H)$_{\rm LDtot}$. 
Nuclear reactions in 
stars and subsequent return of D-poor and metal-rich gas by SN explosions
and stellar winds will reduce deuterium in the ISM by the factor 
$f_d \equiv {\rm (D/H)}_{\rm prim}/{\rm (D/H)}_{\rm LDtot}$. 
Models of \citet{chi02}, 
for example, predict that $f_d \simeq 1.5$ for a wide range of 
assumptions concerning the value of (D/H)$_{\rm prim}$ and amount of mixing 
in stellar interiors. This result can be compared with our results, 
$f_d \leq (27.5^{+2.4}_{-1.9})/(23.1\pm 2.4) \leq 1.19^{+0.16}_{-0.15}$ or
$f_d \leq (26.0^{+1.9}_{-1.7})/(23.1\pm 2.4) \leq 1.12\pm 0.14$. 
These two results differ by $\geq 1.5\sigma$ and $\geq 1.8\sigma$, 
respectively, from the predictions of the \citet{chi02} models.
If, on the other hand, the low value of (D/H)$_{\rm gas} = 8.6\pm 0.8$~ppm 
for the five most distant lines of sight is representative of 
(D/H)$_{\rm LDtot}$ as \citet{heb05} proposes, then $f_d$ would be 
$(27.5^{+2.4}_{-1.9})/(8.6\pm 0.8) = 3.2^{+0.41}_{-0.37}$
or $(26.0^{+1.9}_{-1.7})/(8.6\pm 0.8) = 3.0\pm 0.35$, 
which are far larger values than the \citet{chi02} models predict.
Very recent Galactic chemical evolution models \citep{rom06} can now
accomodate values of $f_d$ as small as $\sim$1.3, consistent with our
determination $f_d < 1.19^{+0.16}_{-0.15}$ or $1.12\pm0.14$, depending on the
adopted value of (D/H)$_{\rm prim}$.
However, our empirical results for $f_d$ are smaller than $f_d = 1.39$, the 
lowest value for a model discussed by \citet{rom06}. We suggest that new
Galactic chemical evolution models should be developed to see whether smaller
values of $f_d$ are consistent with the other empirical abundance and 
theoretical constraints that the models must fit. 

We also call attention to the need for more high quality data 
to better constrain 
(D/H)$_{\rm LDtot}$ and (D/H)$_{\rm prim}$. In particular, we need better
quality observations of the highly
saturated Lyman lines for which measurements of $N$(H~I) may contain
systematic errors of an unknown nature. Also,
the assumptions underlying the Galactic chemical evolution models should be 
reexamined to see whether $f_d$ might be as small as $\sim$1.25 without
violating other constraints.

\acknowledgments 
We thank the {\it FUSE} Science Team for discussions and the successful 
implementation of the {\it FUSE} science program. This
work is supported by NASA grant S-56500-D to NIST and the University of 
Colorado. JLL thanks Uppsala University in Uppsala, Sweden, for hospitality 
where a portion of 
this paper was written and the International Space Science Institute in 
Bern, Switzerland, for support and discussions during workshops on Galactic 
chemical evolution.  We also wish to thank the anonymous referee for his 
critical and insightful comments and suggestions.
This work was supported in part by NSF grant 
AST-9988126 to BTD. This work was also supported by NASA contract NAS-32985 
to Johns Hopkins University.




\begin{deluxetable}{lllll}
\rotate
\tablecaption{PRIMORDIAL D/H: MEASURED AND INFERRED\tablenotemark{a}}
\tablecolumns{5}
\tablewidth{0pt}
\tablehead{
  \colhead{References} &
  \colhead{$\Omega_B h^2$} &
  \colhead{$\eta_{10}$} &
  \colhead{(D/H)$_{\rm prim}$} &
  \colhead{Data Used\tablenotemark{b}}\\
  \colhead{} & \colhead{} & \colhead{} & 
  \colhead{(ppm)} & \colhead{}}
\startdata
\citet{kir03} & $0.0214\pm 0.0020$ & $5.9\pm 0.5$ & $27.8^{+4.4}_{-3.8}$ &
5 quasar absorption line systems\\

\citet{spe03} & $0.0224\pm 0.0009$ & $6.1^{+0.3}_{-0.2}$ & $26.2^{+1.8}_{-2.0}$
& CMB, 2dFGRS, and L$\alpha$ forest data\\

\citet{san06} & $0.0225\pm 0.0010$ & $6.15\pm 0.27$ & $25.9^{+1.9}_{-1.7} $ &
CMB, 2dFGRS, 6 parameter model\\

\citet{cyb03} & $0.0224\pm 0.0009$ & $6.14\pm 0.25$ & $27.5^{+2.4}_{-1.9}$ &
CMB, new nuclear reaction rates\tablenotemark{c}\\ 

\citet{coc04} & $0.0224\pm 0.0009$ & $6.14\pm 0.25$ & $26.0^{+1.9}_{-1.7}$ &
{\it WMAP}, SBBN, new reaction rates\tablenotemark{d}\\

\enddata
\tablenotetext{a}{Quoted uncertainties are $\pm 1\sigma$.}
\tablenotetext{b}{CMB refers to the cosmic microwave background data 
consisting primarily of data from the {\em Wilkinson Microwave Anisotropy 
Probe (WMAP)} but also finer scale CMB experiments. 2dFGRS refers to the power
spectrum of galaxy clustering measured from the final two-degree field galaxy 
redshift survey \citep{san06}. SBBN refers to standard big band 
nucleosynthesis.} 
\tablenotetext{c}{Nuclear reaction rates of \citet{ang99}.}
\tablenotetext{d}{Nuclear reaction rates cited in \citet{coc04}.}

\end{deluxetable}

\begin{deluxetable}{lllllllllllll}
\tabletypesize{\scriptsize}
\rotate
\tablecaption{COMPILATION OF COLUMN DENSITY MEASUREMENTS}
\tablecolumns{11}
\tablewidth{0pt}
\tablehead{
  \colhead{Target} & \colhead{l} & \colhead{b} & \colhead{d} &
    \colhead{$\log N({\rm H~I})$\tablenotemark{a}} &
    \colhead{$\log N({\rm D~I})$\tablenotemark{a}} &  
    \colhead{Satellite} &
    \colhead{Flag\tablenotemark{b}} & 
    \colhead{$\log N({\rm Fe~II})$\tablenotemark{a}} &
    \colhead{$\log N({\rm Si~II})$\tablenotemark{a}} &
    \colhead{Refs.}\\
  \colhead{} & \colhead{(deg)} & \colhead{(deg)} & \colhead{(pc)} &
    \colhead{} & \colhead{} & \colhead{} & 
    \colhead{} & \colhead{} & \colhead{} & \colhead{}\\ 
&\colhead{}}
\startdata
Sirius\tablenotemark{h} & 227 & $-09$ & $2.64\pm 0.02$ & 
$17.60^{+0.14}_{-0.12}$ & $12.81\pm 0.09$ &  HST & LBg &
$11.94^{+0.014}_{-0.016}$ & $12.48^{+0.12}_{-0.08}$ & 42 \\
36 Oph         & 358 & $+07$  & $5.99\pm 0.04$  & $17.85\pm 0.07$ & 
  $13.025\pm 0.01$ & HST & LBg&   $12.65\pm 0.25$  & & 4,27\\
$\epsilon$ Eri & 227 &$-48$&$3.22\pm 0.01$ & $17.875\pm 0.035$ & 
  $13.03\pm 0.03$ & HST & LBg&   $12.26\pm 0.10$  & 
  $12.76\pm 0.14$ & 1,27,28\\
31 Com         & 115 & $+89$ & $94\pm 8$       & $17.884\pm 0.03$ & 
  $13.19\pm 0.025$ & HST & LBg&   $12.55\pm 0.10$  & & 1,27,28\\
HZ 43          &  ~54 & $+84$ & $68\pm 13$      & $17.93\pm 0.03$ & 
  $13.15\pm 0.02$ & FUSE & LBg& $12.17\pm 0.02$  & 
  $12.69\pm 0.06$ & 8,27,28\\
$\epsilon$ Ind & 336 &$-48$&$3.63\pm 0.01$ & $18.00\pm 0.05$ & 
  $13.20\pm 0.027$ &  HST & LBg&   & &   3\\
Procyon        & 214 & $+13$ &$3.50\pm 0.01$   & $18.06\pm 0.05$ &
  $13.26\pm 0.027$ & HST & LBg&   $12.27\pm 0.02$  & & 2,28\\
$\beta$ Cas    & 118 & $-03$& $16.7\pm 0.1$   & $18.132\pm 0.025$ & 
  $13.36\pm 0.03$ & HST & LBg&   $12.36\pm 0.10$  & & 1,27\\
HR 1099        & 185 &$-41$& $29.0\pm 0.7$   & $18.131\pm 0.020$ & 
  $13.295\pm 0.023$ & HST & LBg&  & & 5,27,28\\
G191-B2B       & 156 & $+07$ & $69\pm 15$      & $18.18\pm 0.09$ &
  $13.40\pm 0.035$ & FUSE & LBg& $13.05\pm 0.02$ & 
  $12.55^{+0.15}_{-0.14}$ & 9,27,28\\
$\beta$ CMa\tablenotemark{c}&226 &$-14$& $153\pm 15$ &$18.20^{+0.14}_{-0.20}$ &
  $\geq 13.40$ &  Copernicus&LBg& $13.04\pm 0.04$  & 
  $13.745\pm 0.04$ & 16,29\\
$\sigma$ Gem   & 191 & $+23$ & $37\pm 1$       & $18.201\pm 0.037$ & 
  $13.34\pm 0.05$ &  HST & LBg&   & &   1\\
Capella        & 163 & $+05$  & $12.9\pm 0.1$   & $18.239\pm 0.035$ &
  $13.44\pm 0.02$ & HST & LBg&   $12.49\pm 0.02$  & 
  $13.00\pm 0.05$ & 2,27,28\\
$\beta$ Gem    & 192 & $+23$ & $10.34\pm 0.09$ & $18.261\pm 0.037$ & 
  $13.43\pm 0.05$ & HST & LBg&   $12.42\pm 0.03$  & & 1,27\\
$\alpha$ Tri   & 139 &$-31$& $19.7\pm 0.3$   & $18.327\pm 0.035$ & 
  $13.45\pm 0.05$ & HST & LBg& $12.65\pm 0.20$  & & 1,27\\
$\beta$ Cet    & 111 &$-81$& $29.4\pm 0.7$   & $18.36\pm 0.05$ & 
  $13.70^{+0.08}_{-0.11}$ &  HST & LBg&   & $14.17\pm 0.41$ & 23,28\\
$\lambda$ And  & 110 &$-15$& $25.8\pm 0.5$   & $18.45\pm 0.075$ & 
  $13.68\pm 0.10$ &  HST & LBg &   & &   3\\
Feige 24       & 166 &$-50$& $74\pm 20$      & $18.47\pm 0.03$ & 
  $13.58\pm 0.14$ &  HST  & LBg&   &   $13.54\pm 0.07$ & 10\\
WD 0621-376    & 245 &$-21$& $78\pm 23$      & $18.70\pm 0.15$ &
  $13.85\pm 0.045$ &  FUSE & LBg&    $12.94^{+0.14}_{-0.16}$  & &   11\\
WD 2211-495    & 346 &$-53$& $53\pm 16$      & $18.76\pm 0.15$ &
  $13.94\pm 0.05$ & FUSE & LBg& $13.25\pm 0.075$  & 
  $14.00\pm 0.08$ & 7,30\\
WD 1634-573    & 330 & $-07$& $37\pm 3$       & $18.85\pm 0.06$ &
  $14.05\pm 0.025$ & FUSE & LBg& $13.08\pm 0.065$  & 
  $13.76^{+0.13}_{-0.17}$ & 6,30\\
$\alpha$ Vir   & 316 & $+51$ & $80\pm 6$       & $19.0\pm 0.1$ &
  $14.2^{+0.2}_{-0.1}$ & Copernicus & LBg & $13.16^{+0.01}_{-0.08}$ & 
  $14.16^{+0.01}_{-0.08}$ & 13,32\\
GD 246         &  ~87 &$-45$& $79\pm 24$      & $19.11\pm 0.025$ &
  $14.29\pm 0.045$ & FUSE & LBg& $13.30\pm 0.05$ &
  $14.07^{+0.06}_{-0.04}$ & 12\\
$\lambda$ Sco\tablenotemark{d} & 352 & $-02$ & $216\pm 42$  & $19.23\pm 0.03$&
  $14.11^{+0.09}_{-0.07}$ &  Copernicus & INT & 
  $13.04\pm 0.04$  & $13.42\pm 0.04$ & 19\\
$\beta$ Cen    & 312 & $+01$ & $161\pm 15$     & $19.63\pm 0.10$ &
  $14.7\pm 0.2$ & IUE, Copernicus&
  INT& $13.92^{+0.05}_{-0.04}$ & $14.59\pm 0.01$ & 13,32\\
$\gamma^{2}$ Vel  & 263 & $-08$& $258\pm 35$ & 
  $19.710\pm 0.026$ & $15.05\pm 0.03$ &
  IMAPS & INT & $13.95\pm 0.11$
  & $15.03\pm 0.04$ & 20,26\\
$\alpha$ Cru   & 300 & $+00$ & $98\pm 6$    & $19.85^{+0.07}_{-0.10}$ &
  $14.95\pm 0.05$ & Copernicus&
  INT& $14.00\pm 0.10$ & &13,37\\
BD+28$^{\circ}$4211&~82 &$-19$&$104\pm 18$     & $19.846\pm 0.018$ &
  $14.99\pm 0.025$ & FUSE &INT & 
  $14.10\pm 0.10$ & & 14,38,39\\
Lan 23          & 108 & $-01$ & $122\pm 37$ & $19.89^{+0.25}_{-0.04}$ &
  $15.23\pm 0.065$ & FUSE, EUVE & 
  INT & $14.03\pm 0.065$ & & 12,41\\
$\mu$ Col\tablenotemark{e} & 237 &$-27$& $400^{+100}_{-70}$ & 
  $19.86\pm 0.015$ & $14.7^{+0.3}_{-0.1}$ & HST, 
  Coper.&INT & $14.13\pm 0.02$ & $15.10\pm 0.02$ & 13,35\\
$\zeta$ Pup    & 256 & $-05$& $429\pm 94$     & $19.963\pm 0.026$ &
  $15.11\pm 0.06$ & IMAPS&INT & 
  $14.13^{+0.17}_{-0.07}$ & $15.07^{+0.07}_{-0.02}$ & 20,32\\
TD1~32709 & 233 & +28 & $520\pm 90$ & $20.03\pm0.10$ & $15.30\pm0.05$ & 
  FUSE & INT & $13.95\pm0.10$ &  & 44\\
WD1 1034+001 & 248 & +48 & $155^{+58}_{-43}$ & $20.07\pm0.07$ & 
  $15.40\pm0.07$ & FUSE & INT & $14.10\pm0.10$ &  & 44\\
BD+39$^{\circ}$3226 & ~65 & +29 & $290^{+140}_{-70}$ & 
  $20.08\pm 0.09$ & $15.15\pm0.05$ &   FUSE &  
  INT & $14.15\pm 0.07$ & $14.80\pm 0.20$ & 31,44\\
Feige 110      &  ~74 &$-59$&$179^{+265}_{-67}$&$20.14^{+0.065}_{-0.10}$ &
  $15.47\pm 0.03$ & FUSE, IUE & INT & & &17,38,39\\
$\gamma$ Cas   & 124 & $-02$& $188\pm 20$  & $20.16^{+0.08}_{-0.10}$ &
  $15.15^{+0.04}_{-0.05}$ &  
  Copernicus&INT& $14.08^{+0.11}_{-0.17}$ & & 18,37\\
$\iota$ Ori    & 210 &$-20$& $407\pm 127$    & $20.15^{+0.06}_{-0.07}$ &
  $15.30^{+0.04}_{-0.05}$ &  
  Copernicus & INT & $14.20^{+0.20}_{-0.15}$ & $15.16^{+0.02}_{-0.03}$ & 
  22,32,37,38\\
$\delta$ Ori   & 204 &$-18$& $281\pm 65$     & $20.19\pm 0.03$ &
  $15.06^{+0.06}_{-0.04}$ & 
  IMAPS&INT &  $14.08\pm 0.03$ & & 21,32,40\\
$\theta$ Car\tablenotemark{f} & 290 & $-05$& $135\pm 9$ & $20.28\pm 0.10$ &
  $14.98^{+0.18}_{-0.21}$ & Copernicus&INT& 
  $14.19\pm 0.03$& $14.38^{+0.25}_{-0.08}$ &  15\\
$\epsilon$ Ori & 205 &$-17$& $412\pm 154$ & $20.45^{+0.08}_{-0.10}$ &
  $15.25\pm 0.05$ & Copernicus &
  INT& $14.20\pm 0.10$ & $15.02^{+0.08}_{-0.12}$ & 22,32,37\\
PG 0038+199\tablenotemark{g} & 120 & $-43$& $297^{+164}_{-104}$ & 
  $20.48\pm 0.04$ & $15.76\pm 0.04$ &
  FUSE &INT & $14.42^{+0.03}_{-0.02}$ & & 25\\
LSE 44  & 313 & +13 & $554\pm 66$ & $20.52^{+0.10}_{-0.18}$ & 
  $15.87\pm 0.04$ & FUSE, IUE &
  INT &  &  & 43\\
JL 9           & 323 &$-27$& $590\pm 160$    & $20.78\pm 0.05$ &
  $15.78\pm 0.06$ & FUSE & LDg & 
  $14.69\pm 0.085$ & & 23\\
HD 195965      &  ~86 & $+05$ & $794\pm 200$    & $20.95\pm 0.025$ &
  $15.88\pm 0.07$ & FUSE & LDg &  $14.81\pm 0.01$ & & 24,38\\
LSS 1274       & 277 & $-05$& $580\pm 100$    & $20.98\pm 0.04$ &
  $15.86\pm 0.09$ & FUSE & LDg & 
  $14.81\pm 0.075$  & & 23,39\\
HD 191877      &  ~62 & $-06$ & $2200\pm 550$    & $21.05\pm 0.05$ &
  $15.94^{+0.11}_{-0.06}$ & FUSE & LDg & 
  $14.95\pm 0.02$ & & 24,38\\
HD 90087\tablenotemark{g} & 285 & $-02$ & $2740\pm 800$  & $21.22\pm 0.05$ &
  $16.16\pm 0.06$ & FUSE & LDg  & 
  $15.22\pm 0.02$ & &   41\\

\enddata

\tablenotetext{a}{Quoted uncertainties assumed to be 1$\sigma$ (see
  text). $N$(H~I) is the most recent published value. $N$(D~I) is not 
  listed when not in original paper.\\ 
  $N$(Fe~II) and $N$(Si~II) have been computed using recent, consistent 
  f-values (see \S6.1).}
\tablenotetext{b}{Indicates which lines of sight are used to compute the
  gas-phase Local Bubble (LBg) and gas-phase local disk (LDg) D/H values
  described in\\ 
  the text. INT indicates the intermediate regime, log N(H~I) = 19.2--20.7.}
\tablenotetext{c}{D(Fe) and D(Si) refer to velocity Component C \citep{dup98}.
  N(H~I) and N(D~I) mostly in Component C.}
\tablenotetext{d}{--32 km s$^{-1}$ component only.}
\tablenotetext{e}{D(Fe) and D(Si) refer to Component 1 \citep{how99}, which 
  is also called Component A \citep{shu77}.}
\tablenotetext{f}{Abundances and depletions refer only to the H~I component 
  (Allen et al. 1992).}
\tablenotetext{g}{N(H~I) includes 2N(H$_2$) and (D/H)$_{\rm gas}$ = 
[N(D~I) + N(HD)]/[N(H~I) + 2N(H$_2)$].}
\tablenotetext{h}{The Local Interstellar Cloud component only. For the BC 
component (D/H)$_{\rm gas} = 5^{+11}_{-5}$ ppm.}

\tablerefs{(1) Dring et al.\ 1997. (2) Linsky et al.\ 1995. (3) Wood
  et al.\ 1996. (4) Wood et al.\ 2000. (5) Piskunov et al.\ 1997. (6) Wood
  et al.\ 2002. (7) H\'{e}brard et al.\ 2002. (8) Kruk et al.\ 2002.
  (9) Lemoine et al.\ 2002. (10) Vennes et al.\ 2000. (11) Lehner et al.\
  2002. (12) Oliveira et al.\ 2003. (13) York \& Rogerson 1976.
  (14) Sonneborn et al.\ 2002. (15) Allen et al.\ 1992. (16) Gry et al.\
  1985. (17) Friedman et al.\ 2002. (18) Ferlet et al.\ 1980. (19) York
1983.
  (20) Sonneborn et al.\ 2000. (21) Jenkins et al.\ 1999. (22) Laurent
  et al.\ 1979. (23) Wood et al.\ 2004. (24) Hoopes et al.\ 2003. 
(25) Williger et al.\ 2005.
(26) Fitzpatrick \& Spitzer 1994.
  (27) Redfield \& Linsky 2002. (28) Redfield \& Linsky 2004a.
  (29) Dupin \& Gry 1998. (30) Lehner et al.\ 2003. (31) Bluhm et al.\ 1999.
  (32) van Steenberg \& Shull 1988. (33) Mauche, Raymond, \& Cordova 1988.
  (34) Spitzer \& Fitzpatrick 1993. (35) Howk, Savage, \& Fabian 1999.
  (36) Morton 1978. (37) Jenkins, Savage, \& Spitzer 1986.
  (38) Prochaska et al.\ 2005. (39) H\'ebrard \& Moos 2003.
  (40) Howk et al. 2000. (41) H\'ebrard et al. 2005.
  (42) H\'ebrard et al. 1999. (43) Friedman et al. 2006.
  (44) Oliveira et al. 2006.}
\end{deluxetable}

\begin{deluxetable}{lllllllllllll}
\tabletypesize{\scriptsize}
\rotate
\tablecaption{COMPILATION OF D/H RATIOS AND METAL DEPLETIONS}
\tablecolumns{9}
\tablewidth{0pt}
\tablehead{
  \colhead{Target} & \colhead{l} & \colhead{b} & \colhead{d} &
    \colhead{$\log N({\rm H~I})$\tablenotemark{a}} & 
   \colhead{(D/H)$_{\rm gas}$\tablenotemark{a}} & 
    \colhead{(D/H)$_{\rm gas-LB}$\tablenotemark{a}} & 
    \colhead{D(Fe)\tablenotemark{a}} & 
    \colhead{D(Si)\tablenotemark{a}} \\
  \colhead{} & \colhead{(deg)} & \colhead{(deg)} & \colhead{(pc)} &
    \colhead{} & \colhead{(ppm)} & \colhead{(ppm)} & \colhead{} & \colhead{}\\ 
&\colhead{}}
\startdata
Sirius\tablenotemark{h} & 227 & $-09$ & $2.64\pm 0.02$ & 
$17.60^{+0.14}_{-0.12}$ & $16.2^{+6.4}_{-7.2}$ &   & 
$-1.11^{+0.12}_{-0.14}$ & $-0.63\pm 0.16$ \\
36 Oph         & 358 & $+07$  & $5.99\pm 0.04$  & $17.850\pm 0.075$ & 
  $15\pm 2.5$ &      &    $-0.65\pm 0.27$  & \\
$\epsilon$ Eri & 227 &$-48$&$3.22\pm 0.01$ & $17.875\pm 0.035$ & 
  $14.3\pm 1.6$ &    & $-1.07\pm 0.10$  & 
  $-0.63\pm 0.14$ \\
31 Com         & 115 & $+89$ & $94\pm 8$       & $17.884\pm 0.03$ & 
  $20.2\pm 1.9$ &   &    $-0.78\pm 0.10$  &  \\
HZ 43          &  ~54 & $+84$ & $68\pm 13$      & $17.93\pm 0.03$ & 
  $16.6\pm 1.4$ &   & $-1.21\pm 0.04$  & 
  $-0.75\pm 0.07$ \\
$\epsilon$ Ind & 336 &$-48$ & $3.63\pm 0.01$ & $18.00\pm 0.05$ & 
  $16\pm 2$ &        &    & \\
Procyon        & 214 & $+13$ &$3.50\pm 0.01$   & $18.06\pm 0.05$ &
  $16\pm 2$ &     &    $-1.24\pm 0.05$  & \\
$\beta$ Cas    & 118 & $-03$& $16.7\pm 0.1$   & $18.132\pm 0.025$ & 
  $16.9\pm 1.6$ &  &    $-1.22\pm 0.10$  &\\
HR 1099        & 185 & $-41$ & $29.0\pm 0.7$   & $18.131\pm 0.020$ &
  $14.6\pm 1.0$ &      &   & \\
G191-B2B       & 156 & $+07$ & $69\pm 15$      & $18.18\pm 0.09$ &
  $16.6\pm 4.1$ & & $-0.58\pm 0.09$& 
  $-1.14^{+0.17}_{-0.16}$ \\
$\beta$ CMa\tablenotemark{c} & 226 & $-14$ & $153\pm 15$ & 
  $18.20^{+0.14}_{-0.20}$ & $\geq 16$ &    &  $-0.61^{+0.20}_{-0.14}$  & 
  $+0.04^{+0.20}_{-0.14}$ \\
$\sigma$ Gem   & 191 & $+23$ & $37\pm 1$       & $18.201\pm 0.037$ & 
  $13.8\pm 2.1$ &      &    & \\
Capella        & 163 & $+05$  & $12.9\pm 0.1$   & $18.239\pm 0.035$ &
  $15.9\pm 1.5$&   &  $-1.20\pm 0.04$  & 
  $-0.75\pm 0.06$ \\
$\beta$ Gem    & 192 & $+23$ & $10.34\pm 0.09$ & $18.261\pm 0.037$ & 
  $14.8\pm 2.2$ &  &   $-1.29\pm 0.05$  &\\
$\alpha$ Tri   & 139 &$-31$& $19.7\pm 0.3$   & $18.327\pm 0.035$ & 
  $13.3\pm 2.0$ & & $-1.13\pm 0.20$   &\\
$\beta$ Cet    & 111 &$-81$& $29.4\pm 0.7$   & $18.36\pm 0.05$ &
  $21.9^{+5.2}_{-6.8}$ &      &   & $+0.30\pm 0.41$ \\
$\lambda$ And  & 110 &$-15$& $25.8\pm 0.5$   & $18.45\pm 0.075$ & 
  $17\pm 5$ &      &   & \\
Feige 24       & 166 &$-50$& $74\pm 20$      & $18.47\pm 0.03$ &
  $13\pm 5$ &        &   &   $-0.44\pm 0.08$ \\
WD 0621-376    & 245 &$-21$& $78\pm 23$      & $18.70\pm 0.15$ &
  $14.1\pm 6.0$ &      &    $-1.21\pm 0.20$  &  \\
WD 2211-495    & 346 &$-53$& $53\pm 16$      & $18.76\pm 0.15$ &
  $15.1\pm 6.5$ &  & $-0.96\pm 0.16$  & 
  $-0.27\pm 0.16$ \\
WD 1634-573    & 330 & $-07$& $37\pm 3$       & $18.85\pm 0.06$ &
  $15.8\pm 2.5$ &  &  $-1.22\pm 0.09$  & 
  $-0.60^{+0.14}_{-0.18}$ \\
$\alpha$ Vir   & 316 & $+51$ & $80\pm 6$       & $19.0\pm 0.1$ &
  $15.8^{+10.1}_{-5.8}$ & & $-1.29^{+0.10}_{-0.12}$ &   
  $-0.35^{+0.10}_{-0.12}$ \\
GD 246         &  ~87 &$-45$& $79\pm 24$      & $19.11\pm 0.025$ &
  $15.1\pm 1.9$&   &  $-1.26\pm 0.06$ & $-0.55^{+0.06}_{-0.04}$ \\
$\lambda$ Sco\tablenotemark{d} & 352 & $-02$ & $216\pm 42$  & $19.23\pm 0.03$&
  $~7.6^{+1.8}_{-1.4}$ &     & 
  $-1.64\pm 0.05$  & $-1.32\pm 0.05$ \\
$\beta$ Cen    & 312 & $+01$ & $161\pm 15$     & $19.63\pm 0.10$ &
  $11.7\pm 7.5$ & $~9.5^{+11.9}_{-25.9}$ & 
  $-1.16\pm 0.11$ & $-0.55\pm 0.10$ \\
$\gamma^{2}$ Vel  & 263 & $-08$& $258\pm 35$ & 
  $19.710\pm 0.026$ & 
  $21.8\pm 2.1$ & $24.7\pm 3.2$ & $-1.21\pm 0.11$
  & $-0.19\pm 0.05$\\
$\alpha$ Cru   & 300 & $+00$ & $98\pm 6$    & $19.85^{+0.07}_{-0.10}$ &
  $12.6^{+3.6}_{-2.7}$ & $11.7^{+4.7}_{-3.4}$ & 
  $-1.30^{+0.14}_{-0.12}$ & \\
BD+28$^{\circ}$4211&~82 &$-19$&$104\pm 18$     & $19.846\pm 0.018$ &
  $13.9\pm 1.0$ & $13.4\pm 1.3$ & $-1.20\pm 0.10$ & \\
Lan 23          & 108 & $-01$ & $122\pm 37$ & $19.89^{+0.25}_{-0.04}$ &
  $21.9^{+4.1}_{-17.4}$ & $23.5^{+5.3}_{-23.4}$   &  
  $-1.31^{+0.07}_{-0.25}$ & \\
$\mu$ Col\tablenotemark{e} & 237 &$-27$& $400^{+100}_{-70}$ & 
  $19.86\pm 0.015$ &
  $~6.9^{+6.9}_{-1.8}$ & $~4.5^{+8.2}_{-3.1}$ &  
  $-1.18\pm 0.02$ & $-0.27\pm 0.02$ \\
$\zeta$ Pup    & 256 & $-05$& $429\pm 94$     & $19.963\pm 0.026$ &
  $14.0\pm 2.3$ & $13.7^{+2.7}_{-2.8}$ &  
  $-1.28^{+0.17}_{-0.07}$ & $-0.40^{+0.07}_{-0.03}$ \\
TD1 32709 & 233 & $+28$ & $520\pm 90$ & $20.03\pm 0.10$ & 
  $18.6\pm 5.3$ & $19.1^{+6.7}_{-6.4}$ & $-1.53\pm 0.14$ & \\
WD 1034+001 & 248 & $+48$ & $155^{+58}_{-43}$ & $20.07\pm 0.07$ &
  $21.4\pm 5.3$ & $22.3\pm 6.3$ & $-1.42\pm 0.12$ & \\
BD+39$^{\circ}$3226 & ~65 & +29 & $290^{+140}_{-70}$ & $20.08\pm 0.09$ &
  $11.7\pm 3.1$ & $11.2\pm 3.4$ &  
  $-1.38\pm 0.11$ & $-0.79\pm 0.21$ \\
Feige 110      &  ~74 &$-59$ &$179^{+265}_{-67}$ &$20.14^{+0.065}_{-0.10}$ &
  $21.4^{+5.7}_{-3.8}$ & $22.1^{+6.9}_{-4.4}$ &  & \\
$\gamma$ Cas   & 124 & $-02$& $188\pm 20$  & $20.16^{+0.08}_{-0.10}$ &
  $~9.8^{+2.7}_{-2.3}$ & $~9.1^{+2.9}_{-2.5}$ & 
  $-1.53^{+0.14}_{-0.18}$ & \\
$\iota$ Ori    & 210 &$-20$& $407\pm 127$    & $20.15^{+0.06}_{-0.07}$ &
  $14.1\pm 2.8$ & $13.9\pm 3.1$ & 
  $-1.41^{+0.21}_{-0.16}$ & $-0.51\pm 0.07$ \\
$\delta$ Ori   & 204 &$-18$& $281\pm 65$     & $20.19\pm 0.03$ &
  $~7.4^{+1.2}_{-0.9}$ & $~6.5^{+1.3}_{-1.0}$ &
  $-1.56\pm 0.04$ & \\
$\theta$ Car\tablenotemark{f} & 290 & $-05$& $135\pm 9$ & $20.28\pm 0.10$ &
  $~5.0^{+2.9}_{-3.4}$ & $~4.0^{+3.0}_{-4.5}$ &  
  $-1.54\pm 0.11$& $-1.41^{+0.26}_{-0.12}$ \\
$\epsilon$ Ori & 205 &$-17$& $412\pm 154$ & $20.45^{+0.08}_{-0.10}$ &
  $~6.3^{+1.8}_{-1.5}$ & $~5.8^{+1.8}_{-1.5}$ & 
  $-1.70^{+0.14}_{-0.12}$ & $-0.94^{+0.12}_{-0.14}$ \\
PG 0038+199\tablenotemark{g} & 120 & $-43$& $297^{+164}_{-104}$ & 
  $20.48\pm 0.04$ & 
  $19.1\pm 2.6$& $19.2\pm 2.8$ &  $-1.51^{+0.05}_{-0.04}$ & \\
LSE 44  & 313 & +13 & $554\pm 66$ & $20.52^{+0.10}_{-0.18}$ & 
  $22.4^{+11.7}_{-6.2}$ & $22.7^{+12.8}_{-6.6}$ &   & \\
JL 9           & 323 &$-27$& $590\pm 160$    & $20.78\pm 0.05$ &
  $10.0\pm 1.9$ & $~9.8\pm 2.0$ &  $-1.54\pm 0.10$ & \\
HD 195965      &  ~86 & $+05$ & $794\pm 200$    & $20.95\pm 0.025$ &
  $~8.5\pm 1.6$&$~8.4\pm 1.6$&  $-1.59\pm 0.03$  & \\
LSS 1274       & 277 & $-05$& $580\pm 100$    & $20.98\pm 0.04$ &
  $~7.6\pm 1.9$ & $~7.5\pm 1.9$ &  
  $-1.62\pm 0.08$  & \\
HD 191877      &  ~62 & $-06$ & $2200\pm 550$    & $21.05\pm 0.05$ &
  $~7.8^{+2.4}_{-1.5}$ & $~7.6^{+2.5}_{-1.5}$ &  
  $-1.55\pm 0.05$ & \\
HD 90087\tablenotemark{g} & 285 & $-02$ & $2740\pm 800$  & $21.22\pm 0.05$ &
  $~8.7\pm 1.7$ &$~8.6\pm 1.7$ &   $-1.45\pm 0.05$ & \\
\enddata

\tablenotetext{a}{Quoted uncertainties assumed to be 1$\sigma$ (see
  text). N(H~I) is the most recent published value. N(D~I) is not listed when 
  not\\ 
  in original paper. D/H, D(Fe), and D(Si) are computed using the tablulated 
  value of N(H~I).}
\tablenotetext{b}{Indicates which lines of sight are used to compute the
  gas-phase Local Bubble (LBg) and gas-phase local disk (LDg)\\ 
  D/H values described in the text. 
  INT indicates the intermediate regime, log N(H~I) = 19.2--20.7.}
\tablenotetext{c}{D(Fe) and D(Si) refer to velocity Component C \citep{dup98}.
  N(H~I) and N(D~I) mostly in Component C.}
\tablenotetext{d}{--32 km s$^{-1}$ component only.}
\tablenotetext{e}{D(Fe) and D(Si) refer to Component 1 \citep{how99}, which 
  is also called Component A \citep{shu77}.}
\tablenotetext{f}{Abundances and depletions refer only to the H~I component 
  (Allen et al. 1992).}
\tablenotetext{g}{N(H~I) includes 2N(H$_2$) and (D/H)$_{\rm gas}$ = 
[N(D~I) + N(HD)]/[N(H~I) + 2N(H$_2)$].}
\tablenotetext{h}{The Local Interstellar Cloud component only. For the BC 
component (D/H)$_{\rm gas} = 5^{+11}_{-5}$ ppm.}

\end{deluxetable}

\begin{deluxetable}{llllll}
\tablecaption{ESTIMATES OF (D/H)$_{\rm gas}$ FROM (D/O)$_{\rm gas}\times$ 
(O/H)$_{\rm gas}$} 
\tablecolumns{6}
\tablewidth{0pt}
\tablehead{Reference  & Oxygen  & Number  & Oxygen LOS  & (O/H)$_{\rm gas}$ & 
(D/H)$_{\rm gas}$\tablenotemark{a}\\
  & Data Source & of LOS  & distance (pc) & (ppm) & (ppm)}
\startdata
\citet{heb03} & & & & & \\
~~~cite \citet{mey01} & GHRS, STIS  & 29 & 150--5010 & $343\pm 15$ & 
$13.2\pm 0.8$\\
\citet{and03} & FUSE & 19 & 840--5010 & $408\pm 13$ & $15.7\pm 0.9$\\
~~~include \citet{mey98} & GHRS & 32 & 150--5010 & $377\pm 10$ &$14.5\pm 0.7$\\
~~~include 5 dense LOS & GHRS & 37 & 150--5010 & $362\pm 11$ & $13.9\pm 0.7$\\
\citet{oli05} & FUSE & 11 & 2.6--128 & $345\pm 19$ & $13.2\pm 0.9$\\
\citet{fri03}\tablenotemark{b} & Models 2,8 & -- & 3 & $380\pm 30$ &
 $14.6\pm 1.3$\\
\citet{asp04} & solar\tablenotemark{c} & -- & -- & $458\pm 53$ & 
$17.6\pm 2.1$\\
~~~if 120 ppm in dust & -- & -- & -- & $338\pm 53$ & $13.0\pm 2.0$ \\
\citet{mel04} & solar\tablenotemark{c} & -- & -- & $390\pm 63$ & 
$15.0\pm 2.4$\\
\citet{woo04} & -- & -- & 3.2--94 & -- & $15.6\pm 0.4$\\
\enddata

\tablenotetext{a}{Assuming (D/O)$_{\rm gas} = (3.84\pm 0.16)\times 10^{-2}$ 
(1$\sigma$) \citep{heb03}.}
\tablenotetext{b}{\citet{fri03} do not assign an error to (O/H)$_{\rm gas}$.
We arbitrarily assign an error of 8\% based on the discussion in their paper.}
\tablenotetext{c}{Column 5 lists (O/H)$_{\odot}$, the total abundance ratio 
in the solar photosphere.}
\end{deluxetable}

\begin{deluxetable}{llllllllll}
\tabletypesize{\scriptsize}
\tablecaption{CORRELATION PARAMETERS}
\tablecolumns{9}
\tablewidth{0pt}
\tablehead{
  \colhead{Correlation\tablenotemark{a}} & \colhead{$\log N({\rm H~I})$} &
  \colhead{N} & \colhead{$a$} & 
    \colhead{$\pm 1\sigma$} & \colhead{$b$} & \colhead{$\pm 1\sigma$} & 
    \colhead{Spearman} & \colhead{Significance}}
\startdata
(D/H)$_{\rm gas}$ vs. D(Fe) & all & 38 & 30.00 & ~1.60 & 13.04 & 1.15 & 
0.483 & 0.0021 (2.9$\sigma$)\\

(D/H)$_{\rm gas}$ vs. D(Fe) & HST,FUSE,IMAPS & 29 & 29.14 & ~1.73 & 12.24 & 
1.30 & 0.398 & 0.0324 (2.1$\sigma$)\\

(D/H)$_{\rm gas}$ vs. D(Fe) & $\geq 19.0$ & 24 & 35.04 & ~3.54 & 16.68 & 2.37 &
0.534 & 0.0072 (2.6$\sigma$)\\

(D/H)$_{\rm gas}$ vs. D(Si) & all & 20 & 20.09 & ~0.81 & ~9.72 & 1.05 & 
0.325 & 0.157  (1.4$\sigma$)\\

(D/H)$_{\rm gas}$ vs. D(Si) & $\geq 19.0$ & 11 & 19.27 & ~0.86 & 10.79 & 
0.984 & 0.664 & 0.026  (2.1$\sigma$)\\

(D/Fe)$_{\rm gas}$ vs $\log N({\rm H~I})$ & all & 38 & $-0.59$ & ~0.12 & 
0.081 & 0.006 & 0.516 & 0.0009 (3.1$\sigma$)\\

(D/Fe)$_{\rm gas}$ vs $\log N({\rm H~I})$ & $\geq 19.2$ & 22 & ~1.31 & 
~0.47 & $-0.011$ & 0.023 & 0.0936 & 0.679  (0.4$\sigma$)\\

(D/H)$_{\rm gas-LB}$ vs. $\log T_{01}$ & all & 16 & $-10.66$ & ~4.45 & 
10.54 & 2.22 & 0.564 & 0.023  (2.2$\sigma$)\\

\enddata
\tablenotetext{a}{Correlation equations are of the form $y=a +bx$, where
$a$ is the intercept and $b$ is the slope.}

\end{deluxetable}

\begin{deluxetable}{lllllll}
\tabletypesize{\scriptsize}
\tablecaption{MOLECULAR HYDROGEN TEMPERATURE MEASUREMENTS}
\tablecolumns{7}
\tablewidth{0pt}
\tablehead{
  \colhead{Target} & \colhead{$\log N(\rm H~I)$\tablenotemark{a}} &
    \colhead{$\log N(\rm H_2)$\tablenotemark{a}} &
    \colhead{$\log f(\rm H_2)$\tablenotemark{a}} &
    \colhead{(D/H)$_{\rm gas-LB}$\tablenotemark{a}} & 
    \colhead{$T_{01}$\tablenotemark{a}} &
    \colhead{Refs.}\\
   \colhead{} & \colhead{} & \colhead{} & \colhead{} & \colhead{(ppm)} & 
    \colhead{(K)} & \colhead{}\\
   \colhead{}}

\startdata
$\gamma^2$ Vel\tablenotemark{d} & $19.710\pm 0.026$ & 14.23 & $-5.18$ & 
$24.7\pm 3.2$ & $553^{+1150}_{-292}$ & 1,2\\

BD +28$^{\circ}$4211 & $19.846\pm 0.018$ & $15.13^{+0.20}_{-0.08}$ &
$-4.43^{+0.20}_{-0.08}$ & $13.4\pm 1.3$ & ~$97^{+16}_{-22}$ & 11\\

Lan 23 & $19.89^{+0.25}_{-0.04}$ & 15.11 & $-4.48$ &
$23.5^{+5.3}_{-23.4}$ & $286^{+190}_{-90}$ & 4\\

$\mu$ Col\tablenotemark{d} & $19.86\pm 0.015$ & 15.51 & $-4.09$ &
$~4.5^{+8.2}_{-3.1}$ & ~$98^{+16}_{-13}$ & 2,5\\

$\zeta$ Pup\tablenotemark{d} & $19.963\pm 0.026$ & 14.45 & $-5.21$ &
$13.7^{+2.7}_{-2.8}$ & $382^{+380}_{-158}$ & 1,2\\

TD1 32709\tablenotemark{c} & $20.03\pm 0.10$ & $14.48^{+0.12}_{-0.11}$ &
$-5.25^{+0.15}_{-0.14}$ & $19.1^{+6.7}_{-6.4}$ & $292\pm 83$ & 14\\

WD 1034+001\tablenotemark{c} & $20.07\pm 0.07$ & $15.72^{+0.13}_{-0.12}$ & 
$-4.05^{+0.14}_{-0.13}$ & $22.3\pm 6.3$ & $341\pm 75$ & 14\\

BD+39$^{\circ}$3226 & $20.08\pm 0.09$ & $15.65^{+0.06}_{-0.07}$ &
$-4.13\pm 0.11$ & $11.2\pm 3.4$  & $104\pm 27$ & 14\\

$\iota$ Ori\tablenotemark{d} & $20.16^{+0.06}_{-0.07}$ & 14.69 & $-5.17$ &
$13.6^{+3.1}_{-3.0}$ & $481^{+734}_{-232}$ & 2,6\\

$\delta$ Ori & $20.19\pm 0.03$ & 14.74 & $-5.16$ &
$~6.5^{+1.3}_{-1.0}$ & $598^{+570}_{-243}$ & 7,13\\

$\theta$ Car & $20.28\pm 0.10$ & 15.02 & $-4.96$ &
$~4.0^{+3.0}_{-4.5}$ & ~$96\pm 9$ & 10\\

$\epsilon$ Ori & $20.45^{+0.08}_{-0.10}$ & 16.28 & $-3.89$ &
$~5.8^{+1.8}_{-1.5}$ & ~$89^{+40}_{-24}$ & 6,13\\

PG 0038+199\tablenotemark{b} & $20.48\pm 0.04$ & $19.33\pm 0.02$ & 
$-0.85\pm 0.05$ & $19.2\pm 2.8$ & $139\pm 9$ & 8\\

JL 9\tablenotemark{b} & $20.81\pm 0.05$ & $19.25\pm 0.015$ & $-1.26\pm 0.06$ &
$~9.3\pm 2.0$ & ~$89\pm 3$ & 9\\

LSS 1274\tablenotemark{b} & $20.99\pm 0.04$ & $19.10\pm 0.02$ & 
$-1.59\pm 0.05$ & $~7.3\pm 1.9$ & ~$64\pm 2$ & 9\\

HD 90087\tablenotemark{b} & $21.22\pm 0.05$ & $19.92\pm 0.02$ &
$-1.00\pm 0.05$ & $~8.6\pm 1.7$ & ~$79\pm 2$ & 12\\

\enddata

\tablenotetext{a}{Quoted uncertainties are $1\sigma$.
$f({\rm H_2})=2N({\rm H_2})/[2N({\rm H_2})+N({\rm H~I})]$.\\
$T_{01}=171/ln[9 N({\rm J=0})/N({\rm J=1})]$ K.}

\tablenotetext{b}{$N({\rm H~I})$ includes $2N({\rm H_2})$ and 
(D/H)$_{\rm gas-LB} = 
[N({\rm D~I}) + N({\rm HD})]/[N({\rm H~I}) + 2N({\rm H_2})]$.}

\tablenotetext{c}{$T_{02}$ is given instead of $T_{01}$ which is 
indeterminate.} 

\tablenotetext{d}{Errors in $N({\rm J=0})$ and $N({\rm J=1})$ are 
assumed to be 0.08 ($1\sigma$).}

\tablerefs{(1) Sonneborn et al.\ 2000. (2) Savage et al.\ 1977.
  (3) Bluhm et al. \ 1999.  (4) Oliveira et al.\ 2003. 
  (5) York \& Rogerson 1976.  (6) Laurent et al.\ 1979.
  (7) Jenkins et al.\ 1999.   (8) Williger et al.\ 2005.  
  (9) Wood et al.\ 2004.  (10) Allen et al. 1992. (11) Sonneborn et al.\ 2002.
  (12) H\'ebrard et al. 2005. (13) Jenkins et al. 2000b.
  (14) Oliveira et al. 2006.}
\end{deluxetable}

\clearpage
\begin{figure*}
\plotone{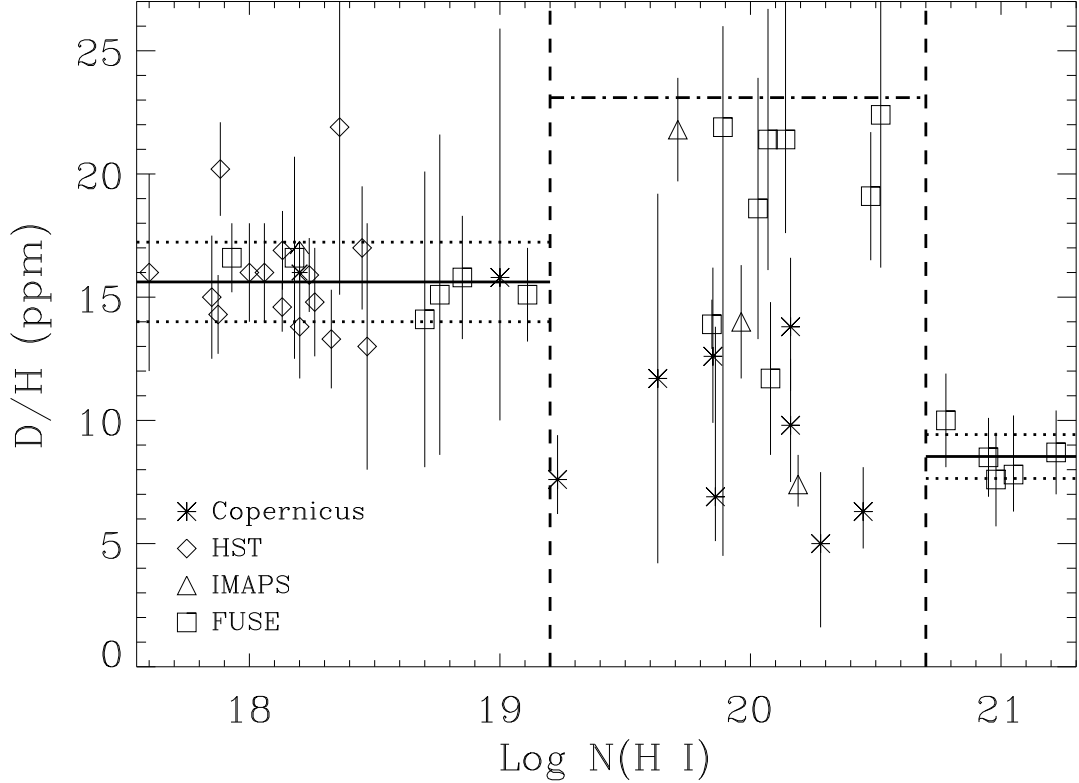}\vspace*{0.5cm}
\caption{Plot of the gas-phase D/H values (not corrected for the Local Bubble
foreground) vs. the hydrogen column density for the 47 lines of sight listed in
Table~3. The symbols for each data point indicate the spacecraft that observed
the line of sight. Error bars are $\pm1~\sigma$. Vertical dashed lines indicate
lines of sight inside the Local Bubble [$\log N({\rm H~I}) < 19.2$ cm$^{-2}$] 
and the
intermediate region [$\log N({\rm H~I})$ between 19.2 and 20.7].
The solid horizontal lines indicate the mean values of
(D/H)$_{\rm gas}$ for the low and high $N({\rm H~I})$ regions, 
and the dotted horizontal lines indicate the
$\pm 1~\sigma$ about the mean value. The dash-dot horizontal line is the 
mean value of (D/H)$_{\rm gas-LB}$ for the highest five points 
in the intermediate region ($\gamma^2$~Vel, Lan~23, WD~1034+001, Feige~110, 
and LSE~44) after 
subtracting the foreground Local Bubble contributions to the hydrogen and 
deuterium column densities along the lines of sight. See discussion in \S 7.} 
\end{figure*}

\clearpage
\begin{figure*}
\plotone{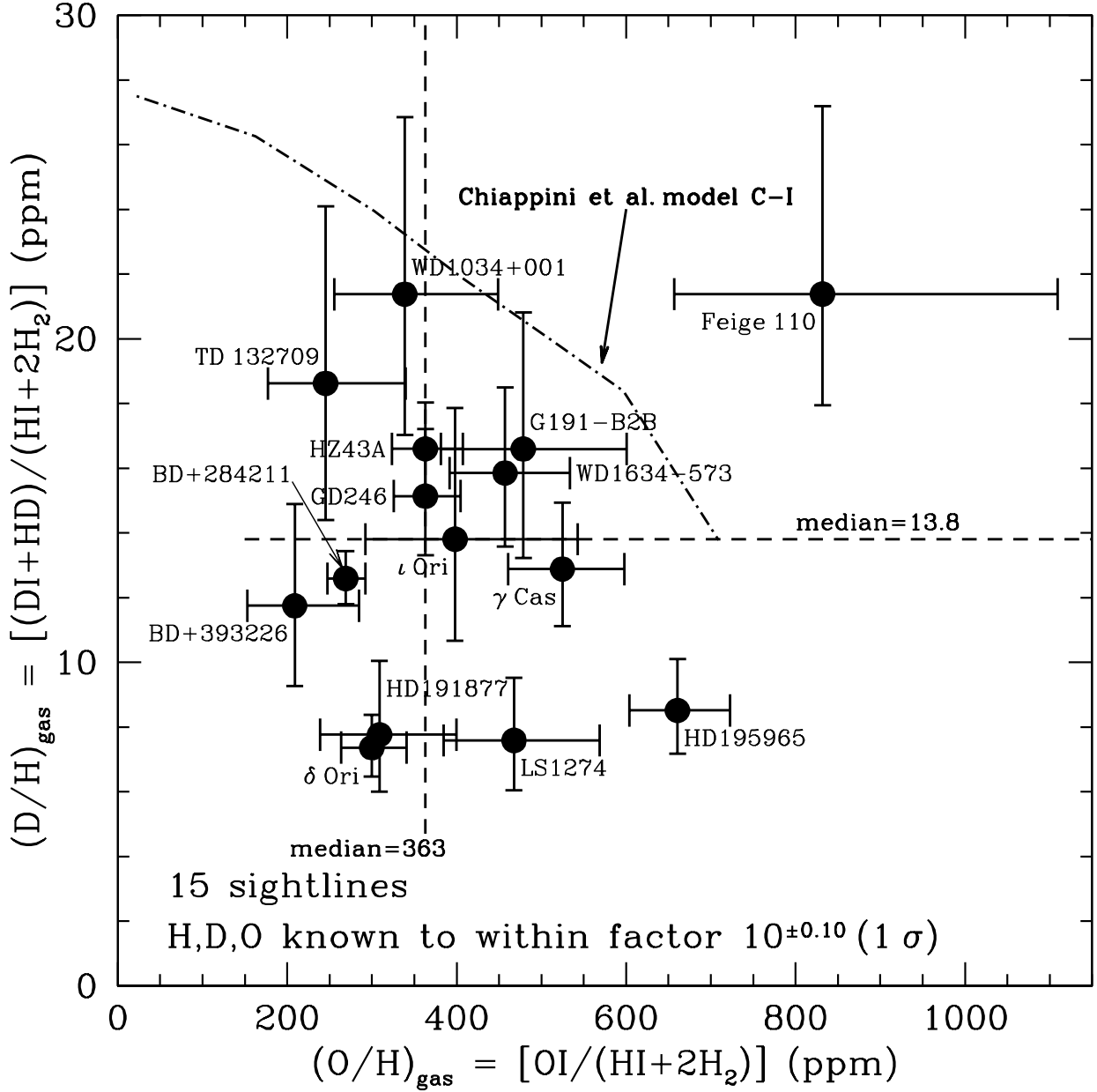}\vspace*{0.5cm}
\caption{\label{fig:DvsO}
         \footnotesize
	 (D/H)$_{\rm gas}$ vs.\ (O/H)$_{\rm gas}$ with $1\sigma$ errors
	 (see text).  The data do not show anticorrelation of
	 (D/H) with (O/H), as would result from variable astration or
	 infall, such as model C--I from Chiappini et al.\ (2002).
	 The full range of D/H is seen for sightlines with
	 (O/H)$_{\rm gas}$ between 300 and 390 ppm.
	 Depletion of D into interstellar dust appears to be the
	 most viable explanation for the observed variations
	 in (D/H)$_{\rm gas}$.
	 \newline
	 Sightlines: 
	 WD~1634-573 (Wood et al.\ 2002);
	 GD~246 (Oliveira et al.\ 2003);
	 BD+284211 (Sonneborn et al.\ 2002; Spitzer et al.\ 1974);
	 Feige 110 (Friedmann et al.\ 2002; H\'ebrard et al.\ 2005);
	 $\gamma$~Cas (Ferlet et al.\ 1980; Meyer et al.\ 1988; Meyer 2001);
	 $\delta$~Ori (Jenkins et al.\ 1999; Meyer et al.\ 1998; Meyer 2001);
	 $\iota$~Ori (Laurent et al.\ 1979; Meyer et al.\ 1998; Meyer 2001);
	 HD 195965 and HD 191877 (Hoopes et al.\ 2003);
	 HZ~43A: (Kruk et al.\ 2002);
	 G~191-B2B (Lemoine et al.\ 2002);
	 LSS~1274 (Wood et al.\ 2004); H\'ebrard \& Moos (2003).
	 }
\end{figure*}

\clearpage
\begin{figure*}
\plotone{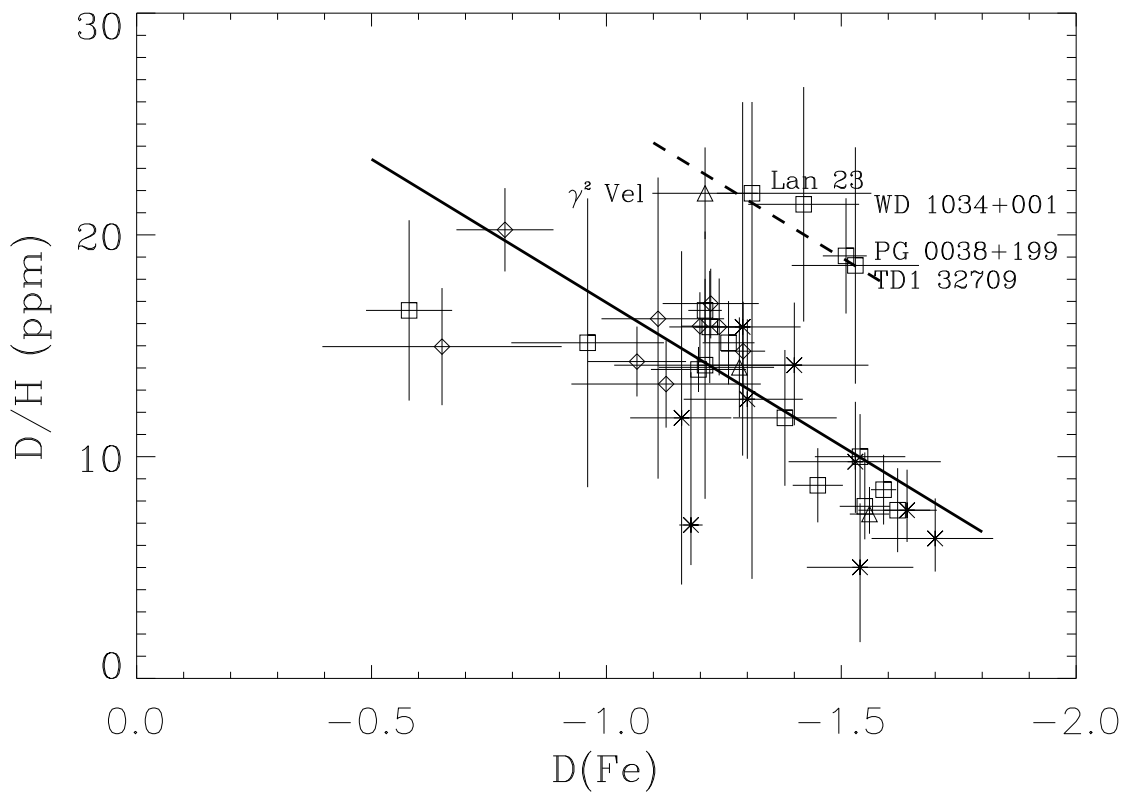}\vspace*{0.5cm}
\caption{Plot of gas-phase D/H values toward stars vs. the depletion of iron 
for the 38 lines of sight listed in Table~3 (ignoring upper limits). 
The symbols are the same as in Figure 1, and the error bars are $\pm1~\sigma$.
The solid line is the least-squares linear fit to the 
data weighted by the inverse errors. See \S6 for a discussion of the 
error analysis technique. Five lines of sight with
(D/H)$_{\rm gas}$ well above the linear fit are identified. The dashed line
has the same slope as the solid line but is displaced upward by 8.5~ppm.
See discussion in \S 6.3.4.}
\end{figure*}

\clearpage
\begin{figure*}
\plotone{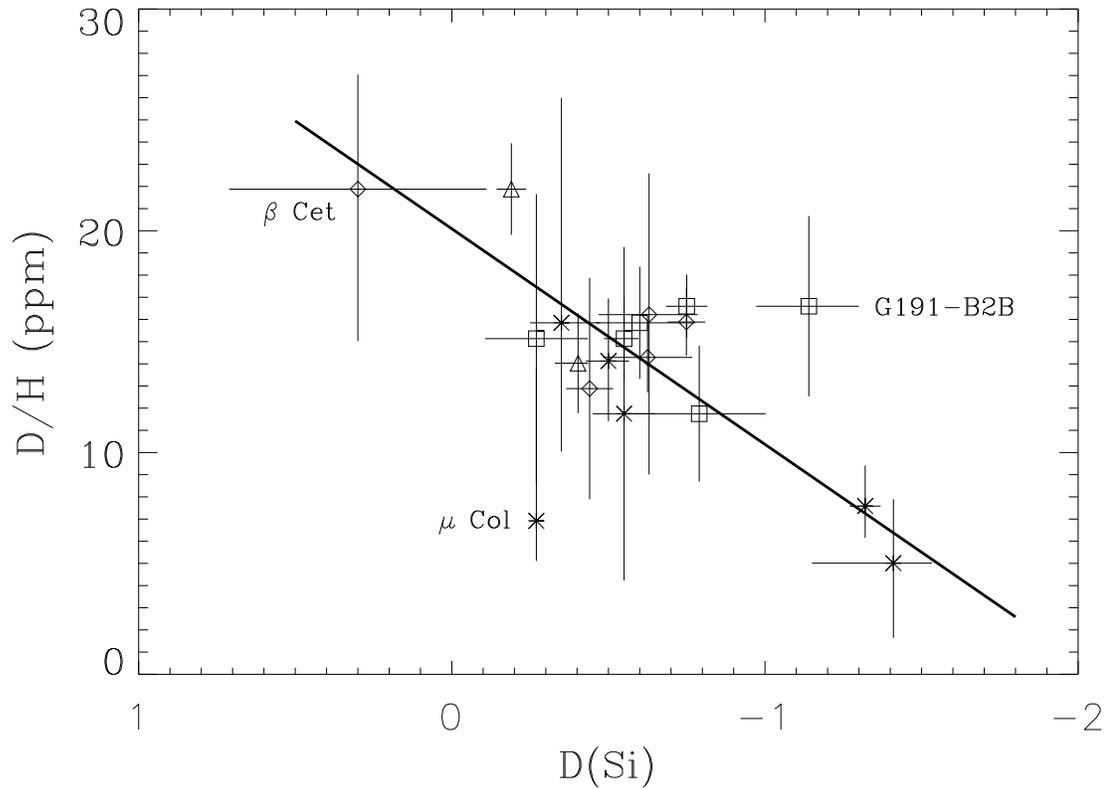}\vspace*{0.5cm}
\caption{Plot of gas-phase D/H values toward stars vs. the depletion of 
silicon for the 20 lines of sight (ignoring upper limits) listed in Table~3. 
The symbols are the same as in Figure 1, and the error bars are $\pm 
1~\sigma$. The solid line is the least-squares linear 
fit to the data (except for upper limits) weighted by the inverse errors.
Lines of sight to three stars are identified.}
\end{figure*}

\clearpage
\begin{figure*}
\plotone{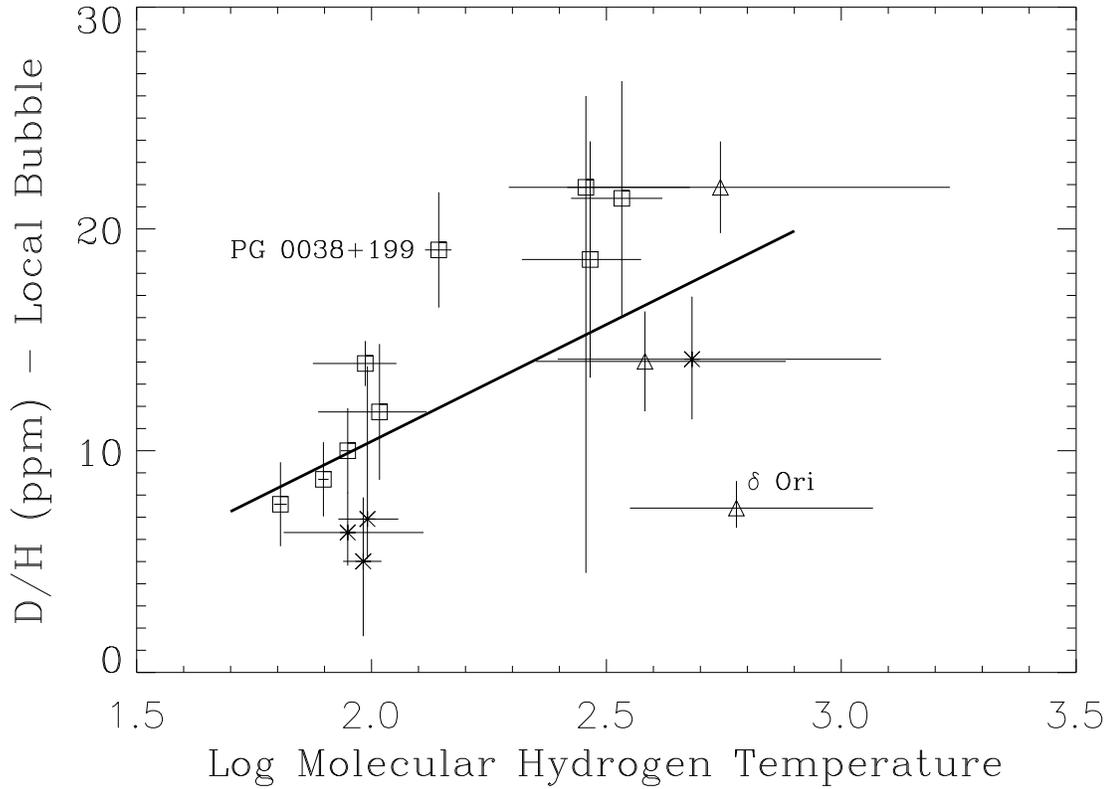}\vspace*{0.5cm}
\caption{Plot of gas-phase D/H values toward stars (after correction for the 
Local Bubble foreground) vs. $T_{01}$, the excitation
temperature of molecular hydrogen between the J=0 and J=1 rotational levels. 
The solid line is the least-squares linear fit to the 16 data 
points (see Table~6) weighted by the inverse errors. 
Two outliers are identified.}

\end{figure*}

\clearpage
\begin{figure*}
\plotone{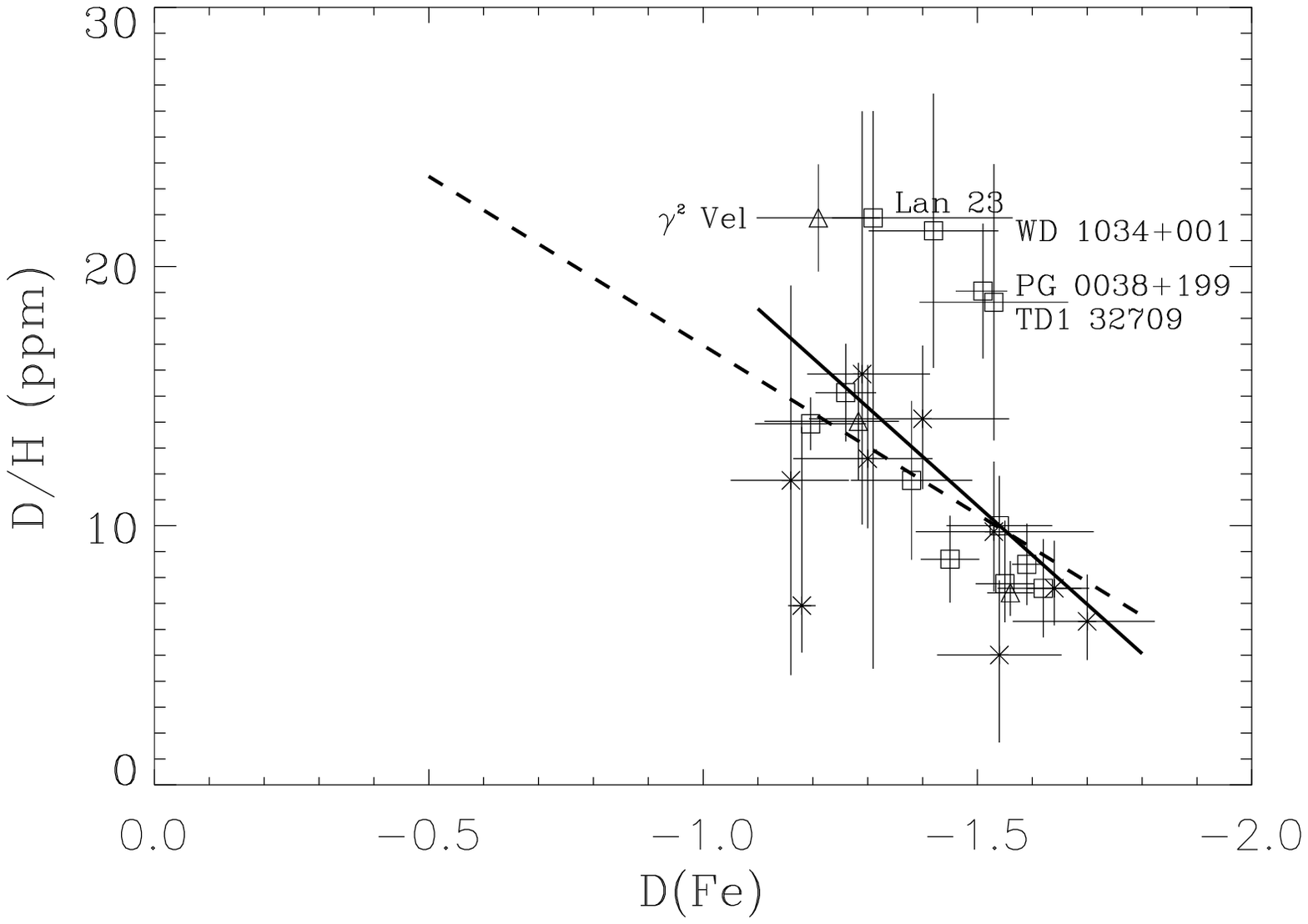}\vspace*{0.5cm}
\caption{Plot of gas-phase D/H values toward stars vs. the depletion of iron 
for the 24 lines of sight listed in Table~3 with $\log N({\rm H~I})\geq 19.0$. 
The symbols are the same as in Figure~1, and the error bars are $\pm1~\sigma$.
The solid line is the least-squares linear fit to the 
data weighted by the inverse errors. The dashed line is the
fit to the 38 lines of sight covering the full range of $\log N({\rm H~I})$ as 
shown in Figure~3.}
\end{figure*}

\clearpage
\begin{figure*}
\plotone{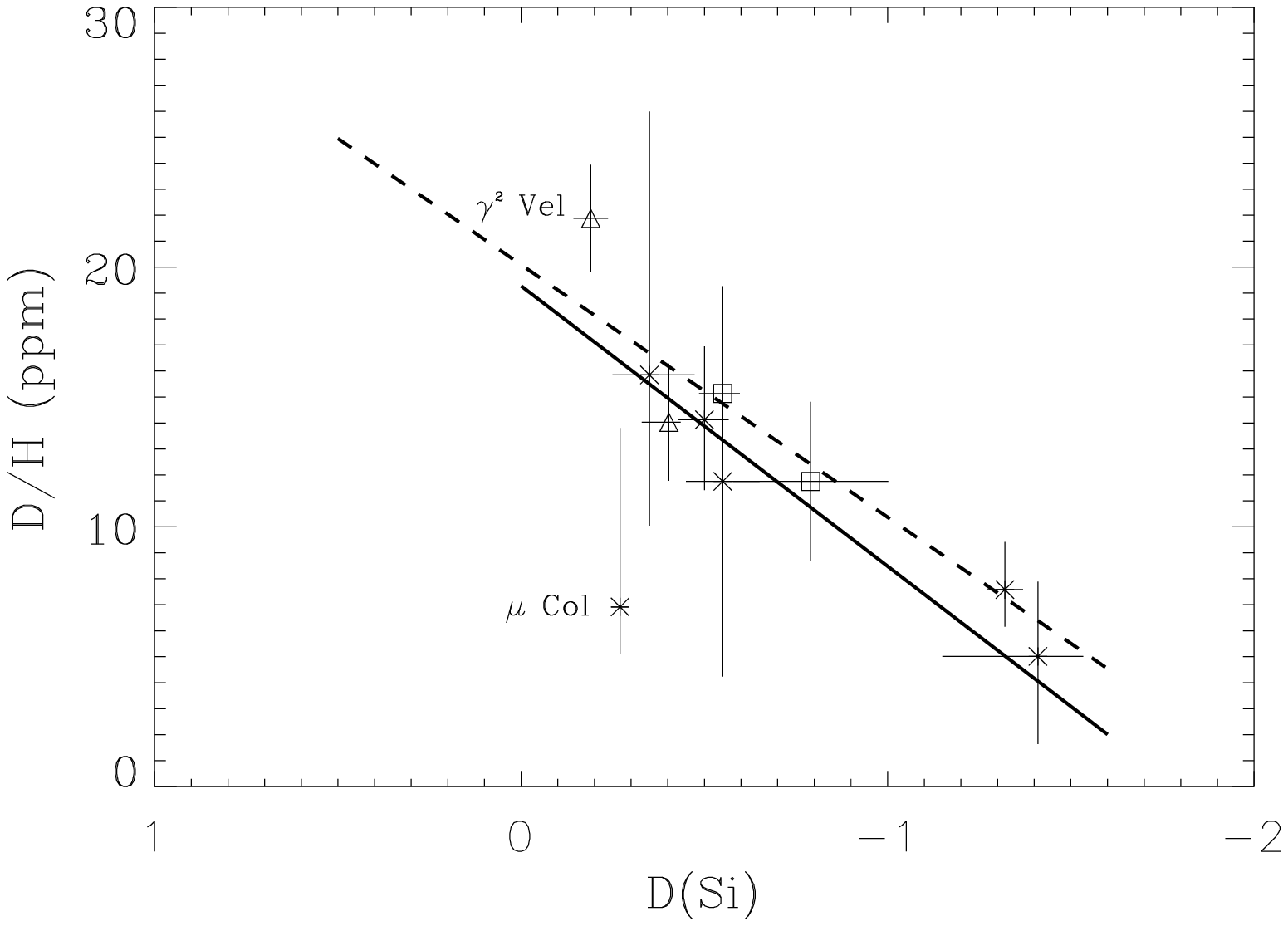}\vspace*{0.5cm}
\caption{Plot of gas-phase D/H values toward stars vs. the depletion of 
silicon for the 11 lines of sight listed in Table~3 with 
$\log N({\rm H~I})\geq 19.0$.
The symbols are the same as in Figure~1, and the error bars are $\pm 
1~\sigma$. The solid line is the least-squares linear 
fit to the data weighted by the inverse errors. The dashed line
is the fit to the 20 lines of sight covering the full range of 
$\log N({\rm H~I})$ as shown in Figure~4.}
\end{figure*} 

\clearpage
\begin{figure*}
\plotone{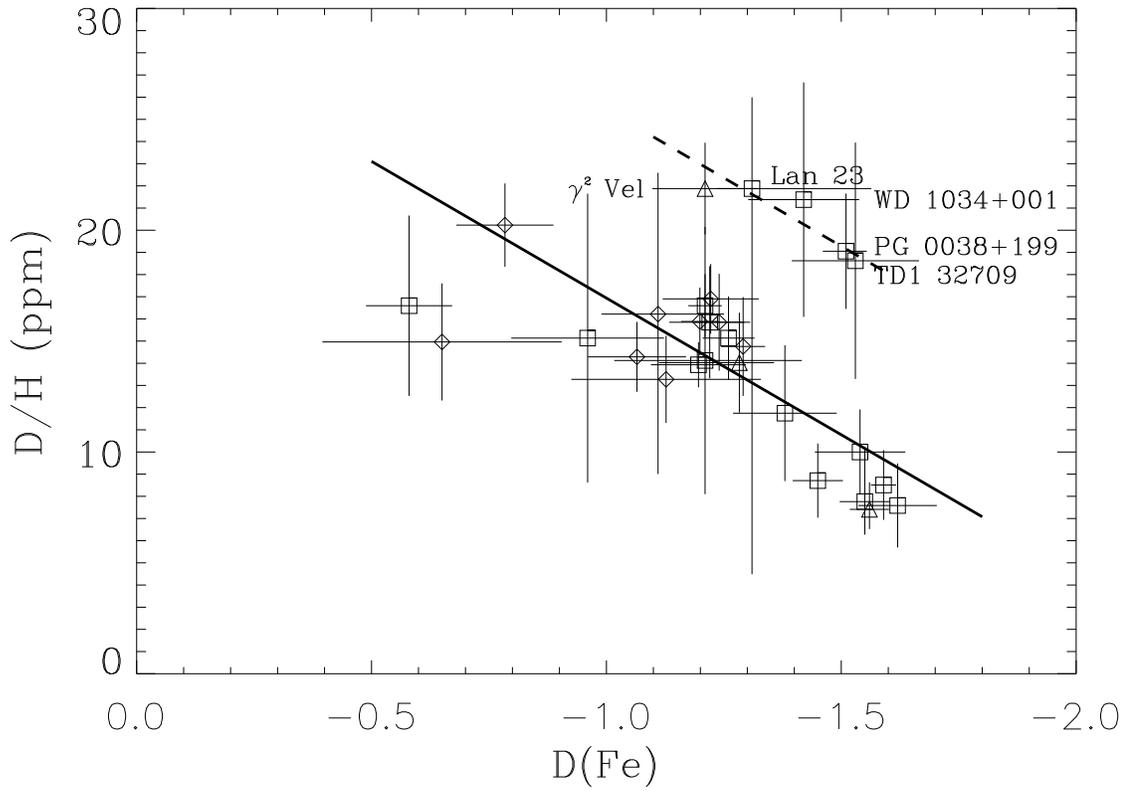}\vspace*{0.5cm}
\caption{Same as Figure 3 except using data only from {\em STIS}, {\em GHRS}, 
{\em IMAPS}, and {\em FUSE}.}
\end{figure*}

\clearpage
\begin{figure*}
\plotone{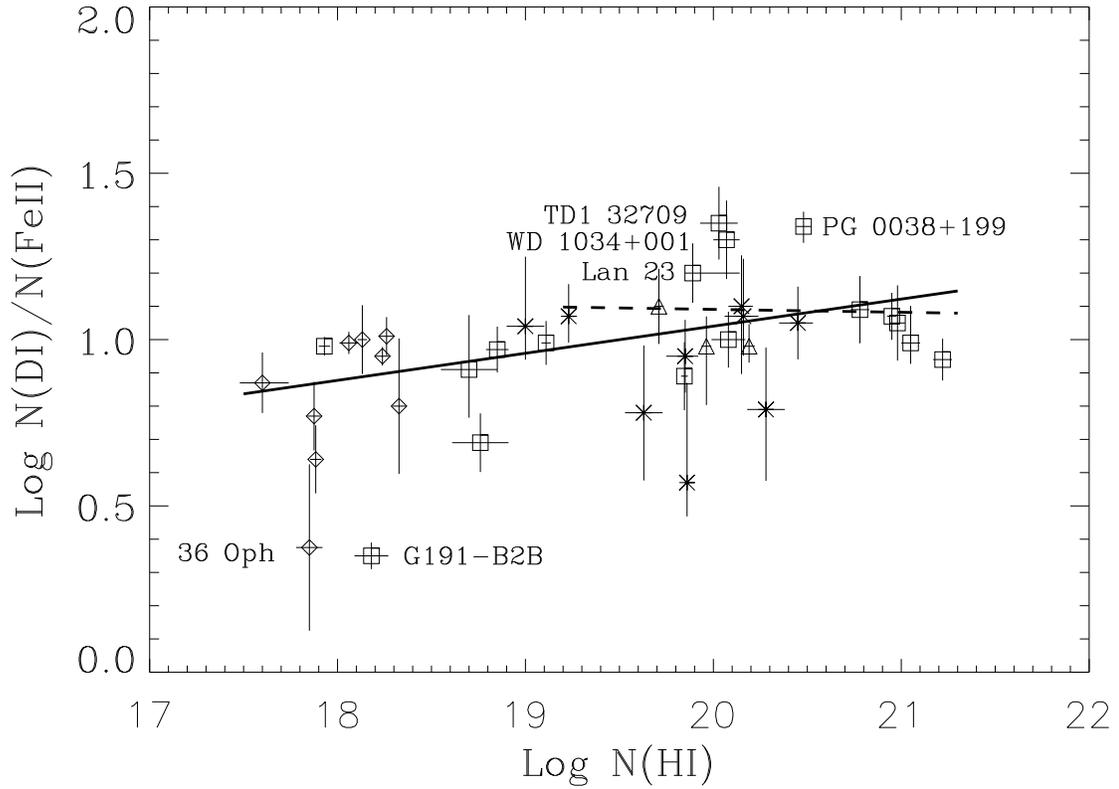}
\caption{Column density ratios $\log [N({\rm D~I})/N({\rm Fe~II})]$ vs. 
$\log N({\rm H~I})$ for all 
lines of sight with symbol coding the same as in Figure~1. The solid line is a 
weighted least-squares fit to all of the data points, and the dashed line 
is a similar fit to only the data beyond the Local Bubble 
($\log N({\rm H~I}) > 19.2$). 
}
\end{figure*}


\begin{thebibliography}{}

\bibitem[Allen, Jenkins, \& Snow(1992)]{all92} 
Allen, M. M., Jenkins, E. B., \& Snow, T. P. 1992, ApJS,
83, 261 

\bibitem[Allende Prieto et al.(2001)]{all01} Allende Prieto, C., Lambert, 
D. L., \& Asplund, M. 2001, \apj, 556, L63

\bibitem[Andr\'e et al.(2003)]{and03} Andr\'e, M. K. et al. 2003, \apj, 591,
1000

\bibitem[Anderson et al.(2004)]{and04} Anderson, K. L., Shull, J. M., \& 
Tumlinson, J. 2004, Bull. AAS, 36, 772          

\bibitem[Anders \& Grevesse(1989)]{and89} Anders, E. \& Grevesse, N. 1989,
Geochim. Cosmochim. Acta 53, 197

\bibitem[Angulo et al.(1999)]{ang99} Angulo, C. et al. 1999, Nuclear Physics 
A, 656, 3

\bibitem[Asplund et al.(2004)]{asp04} Asplund, M., Grevesse, N., Sauval, 
A. J., Allende Prieto, C., \& Kiselman, D. 2004, \aap, 417, 751

\bibitem[Asplund et al.(2005)]{asp05} Asplund, M., Grevesse, N. \& Sauval, 
A. J. 2005, in Cosmic Abundances as Records of Stellar Evolution and 
Nucleosynthesis, ed. F. N. Bash \& T. G. Barnes 
(San Francisco: ASP Conf. Series), 336, 25 

\bibitem[Bahcall, Serenelli, \& Basu(2005)]{bah05} Bahcall, J. N., Serenelli, 
A. M., \& Basu, S. 2005, \apj, 621, L85

\bibitem[Bergh\"ofer \& Breitschwerdt(2002)]{ber02} Bergh\"ofer, T. W. \& 
Breitschwerdt, D. 2002, \aap, 390, 299


\bibitem[Bluhn et al.(1999)]{blu99} Bluhm, H., Marggraf, O., de Boer, K. S., 
Richter, P., \& Heber, U. 1999, \aap, 352, 287

\bibitem[Boesgaard \& Steigman(1985)]{boe85} Boesgaard, A. M., \& Steigman, G. 
1985, \araa, 23, 319

\bibitem[Burles, Nollett, \& Turner(2001)]{bur01} 
Burles, S., Nollett, K. M., \& Turner, M. S. 2001, ApJ,
552, L1 

\bibitem[Cartledge et al.(2004)]{car04} Cartledge, S. I. B., Lauroesch, J. T., 
Meyer, D. M., \& Sofia, U. J. 2004, \apj, 613, 1037

\bibitem[Chiappini, Renda, \& Matteucci(2002)]{chi02} Chiappini, C., Renda, 
A., \& Matteucci, F. 2002, \aap, 395, 789

\bibitem[Coc et al.(2004)]{coc04} Coc, A., Vangioni-Flan, E., Descouvemont, 
P., Adahchour, A., \& Angulo, C. 2004, \apj, 600, 544 

\bibitem[Collins et al.(2003)]{col03} Collins, J. A., Shull, J. M., \& Giroux, 
M. L. 2003, \apj, 585, 336


\bibitem[Cox \& Reynolds(1987)]{cox87} Cox, D. P., \& Reynolds, R. J. 1987, 
\araa, 25, 303

\bibitem[Cyburt, Fields, \& Olive(2003)]{cyb03} Cyburt, R. H., Fields, B. D.,
\& Olive, K. A. 2003, Physics Letters B, 567, 227 


\bibitem[de Avillez \& Mac Low(2002)]{dea02} de Avillez, M. A. \& Mac Low, 
M.-M. 2002, \apj, 581, 1047

\bibitem[Draine(2003)]{dra03} Draine, B. T. 2003, \araa, 41, 241

\bibitem[Draine(2004)]{dra04} Draine, B. T. 2004, in Origin and Evolution of 
the Elements, ed. A. McWilliam \& M. Rauch (Cambridge: Cambridge Univ. Press),
320

\bibitem[Draine(2006)]{dra06} Draine, B. T. 2006, in Astrophysics in the Far
Ultraviolet: Five Years of Discovery with FUSE, ed: G. Sonneborn, 
B. G. Andersson \& Warren Moos (San Francisco: ASP), in press

\bibitem[Dupin \& Gry(1998)]{dup98} Dupin, O. \& Gry, C. 1998, \aap, 335, 661

\bibitem[Dring et al.(1997)]{dri97}
Dring, A. R., Linsky, J., Murthy, J., Henry, R. C., Moos, W., Vidal-Madjar,
  A., Audouze, J., \& Landsman, W. 1997, ApJ, 488, 760


\bibitem[Epstein, Lattimer, \& Schramm (1976)]{eps76} Epstein, R. I., 
Lattimore, J. M., \& Schramm, D. N. 1976, Nature, 263, 198

\bibitem[Ferlet et al.(1980)]{fer80}
Ferlet, R., Vidal-Madjar, A., Laurent, C., \& York, D. G. 1980, ApJ, 242, 576

\bibitem[Fitzpatrick \& Spitzer(1994)]{fit94} Fitzpatrick, E. L., \& Spitzer
Jr., L. 1994, \apj, 427, 232

\bibitem[Friedman et al.(2002)]{fri02}
Friedman, S. D., et al. 2002, ApJS, 140, 37

\bibitem[Friedman et al.(2006)]{fri06} Friedman, S. D., H\'ebrard, G., 
Tripp, T. M., Chayer, P., \& Sembach, K. R. 2006, \apj, 638, 847

\bibitem[Frisch et al.(1999)]{fri99} Frisch, P.~C., et al. 1999, \apj, 525, 492

\bibitem[Frisch \& Slavin(2003)]{fri03} Frisch, P.~C. \& Slavin, J.~D.
2003, \apj, 594, 844

\bibitem[Geiss \& Gloeckler(1998)]{gei98} Geiss, J. \& Gloeckler, G. 1998,
\ssr, 84, 239

\bibitem[Geiss, Gloeckler, \& Charbonnel(2002)]{gei02} Geiss, J., Gloeckler, 
G., \& Charbonnel, C. 2002, \apj, 578, 862

\bibitem[Gillmon et al.(2006)]{gil06} Gillmon, K., Shull, J. M., Tumlinson, J.,
\& Danforth, C. 2006, \apj, 636, 891

\bibitem[Gry, York, \& Vidal-Madjar(1985)]{gry85} 
Gry, C., York, D. G., \& Vidal-Madjar, A. 1985, ApJ, 296, 593 

\bibitem[H\'ebrard et al.(1999)]{heb99} H\'ebrard, G., Mallouris, C., 
Koester, D., Lemoine, M., Vidal-Madjar, A., \& York, D. 1999, \aap, 350, 643 

\bibitem[H\'{e}brard et al.(2002)]{heb02}
H\'{e}brard, G., et al. 2002, ApJS, 140, 103

\bibitem[H\'ebrard et al.(2005)]{heb05} H\'ebrard, G., Tripp, T. M., 
Chayer, P., Friedman, S. D., Dupuis, J.,
Sonnentrucker, P., Williger, G. M., \& Moos, H. W. 2005, \apj, 635, 1136

\bibitem[H\'ebrard \& Moos(2003)]{heb03} H\'ebrard, G. \& Moos, H. W. 2003, 
\apj, 599, 297

\bibitem[Hoopes et al.(2003)]{hoo03} Hoopes, C. G., Sembach, K. R., H\'ebrard,
G., Moos, H. W., \& Knauth, D. C. 2003, \apj, 586, 1094 

\bibitem[Howk et al.(1999)]{how99} Howk, J. C., Savage, B. D., \& Fabian, D. 
1999, \apj, 525, 253

\bibitem[Howk et al.(2000)]{how00} Howk, J. C., Sembach, K. R., Roth, K. C., \&
Kruk, J. W. 2000, \apj, 544, 867

\bibitem[Jenkins(1996)]{jen96} Jenkins, E. B. 1996, in Cosmic Abundances, ed.
S. S. Holt \& G. Sonneborn (San Francisco: ASP Conf. Series), 99, 90

\bibitem[Jenkins(2004)]{jen04} Jenkins, E. B. 2004, in Origin and Evolution of 
the Elements, ed. A. McWilliam \& M. Rauch (Cambridge: Cambridge Univ. Press), 
336

\bibitem[Jenkins et al.(2000a)]{jen00a} Jenkins, E. B., Gry, C., \& Dupin, O.
2000a, \aap, 354, 253

\bibitem[Jenkins et al.(2000b)]{jen00b} Jenkins, E. B., Wozniak, P. R., 
Sofia, U. J., Sonneborn, G., \& Tripp, T. M. 2000b, \apj, 538, 275 

\bibitem[Jenkins et al.(1999)]{jen99} Jenkins, E. B., Tripp, T. M., Wozniak, 
P. R., Sofia, U. J., \& Sonneborn, G. 1999, ApJ, 520, 182

\bibitem[Jenkins et al.(1986)]{jen86} Jenkins, E. B., Savage, B. D., 
\& Spitzer Jr., L. 1986, \apj, 301, 355

\bibitem[Jones(2004)]{jon04} Jones, A. P. 2004, JGR, 105, A5, 10257

\bibitem[Jones et al.(1994)]{jon94} Jones, A. P., Tielens, A. G. G. M., 
Hollenbach, D. J., \& McKee, C. F. 1994, \apj, 433, 797

\bibitem[Jura(1982)]{jur82} Jura, M. 1982, in Advances in UV Astronomy: 4 Years
of IUE Research, ed. Y. Kondo, J. M. Mead, \& Chapman, R. D. (NASA CP 2238;
Greenbelt MD: NASA), 54 

\bibitem[Keller, Messenger, \& Bradley(2000)]{kel00} Keller, L. P., Messenger, 
S., \& Bradley, J. P. 2000, \jgr, 105, 10397

\bibitem[Kirkman et al.(2003)]{kir03} Kirkman, D., Tytler, D., Suzuki, N., 
O'Meara, J., \& Lubin, D. 2003, \apjs, 149, 1

\bibitem[Kruk et al.(2002)]{kru02}
Kruk, J. W., et al. 2002, ApJS, 140, 19



\bibitem[Lallement et al.(2003)]{lal03} Lallement, R., Welsh, B. Y., Vergely, 
J. L., Crifo, F., \& Sfeir, D. 2003, \aap, 411, 447

\bibitem[Laurent, Vidal-Madjar, \& York(1979)]{lau79} Laurent, C., 
Vidal-Madjar, A., \& York, D. G. 1979, \apj, 229, 923


\bibitem[Lehner et al.(2002)]{leh02}
Lehner, N., Gry, C., Sembach, K. R., H\'{e}brard, G., Chayer, P., Moos, H. W.,
  Howk, J. C., \& D\'{e}sert, J. -M. 2002, ApJS, 140, 81

\bibitem[Lehner et al.(2003)]{leh03} Lehner, N., Jenkins, E. B., Gry, C., 
Moos, H. W., Chayer, P., \& Lacour, S. 2003, \apj, 595, 858 

\bibitem[Lehner et al.(2004)]{leh04} Lehner, N., Wakker, B. P., 
\& Savage, B. D. 2004, \apj, 615, 767

\bibitem[Lellouch et al.(2001)]{lel01} Lellouch, E., B\'ezard, B., 
Fouchet, T., Feuchtgruber, H., Encrenaz, T., \& de Graauw, T. 2001,
\aap, 670, 610 

\bibitem[Lemoine et al.(2002)]{lem02}
Lemoine, M., et al. 2002, ApJS, 140, 67


\bibitem[Li \& Draine(2001)]{lid01}
Li, A., \& Draine, B. T. 2001, \apj, 554, 778

\bibitem[Linsky(1998)]{lin98} Linsky, J. L. 1998, Space Sci. Rev. 84, 285

\bibitem[Linsky(2003)]{lin03} Linsky, J. L. 2003, Space Sci. Rev. 106, 49

\bibitem[Linsky(2006)]{lin06} Linsky, J. L. 2006, 
in Astrophysics in the Far Ultraviolet: Five Years of Discovery with FUSE, 
ed. G. Sonneborn, B. G. Andersson \& Warren Moos (San Francisco: ASP), 
in press

\bibitem[Linsky et al.(1995)]{lin95}
Linsky, J. L., Diplas, A., Wood, B. E., Brown, A., Ayres, T. R., \&
Savage, B. D. 1995, ApJ, 451, 335


\bibitem[Lockman(2002)]{loc02} Lockman, F. J. 2002, \apj, 580, L47

\bibitem[Lockman, Jahoda, \& McCammon(1986)]{loc86} Lockman, F. J., Jahoda, 
K., \& McCammon, D. 1986, \apj, 302, 432

\bibitem[Lyu \& Bruhweiler(1996)]{lyu96} Lyu, C. H. \& Bruhweiler, F. C. 
1996, \apj, 459, 216

\bibitem[Markwick et al.(2001)]{mar01} Markwick, A. J., Charnley, S. B., 
\& Millar, T. J. 2001, \aap, 376, 1054

\bibitem[Ma\'iz-Apell\'aniz(2001)]{mai01} Ma\'iz-Apell\'aniz, J. 2001, \apj, 
560, L83

\bibitem[Mauche et al.(1988)]{mau88} Mauche, C. W., Raymond, J. C., \& 
Cordova, F. A. 1988, \apj, 335, 829

\bibitem[Mel\'endez(2004)]{mel04} Mel\'endez, J. 2004, \apj, 615, 1042

\bibitem[Meyer(2001)]{mey01} Meyer, D. M. 2001, in Gaseous Matter in Galaxies 
and Intergalactic Space, ed. R. Ferlet et al. (Paris: Editions Fronti\'ers),
135

\bibitem[Meyer et al.(1998)]{mey98} Meyer, D. M., Jura, M., \& Cardelli, J. A.
1998, \apj, 493, 222

\bibitem[Moos et al.(2000)]{moo00} Moos, H. W. et al. 2000, \apj, 538, L1

\bibitem[Moos et al.(2002)]{moo02} Moos, H. W. et al. 2002, \apjs, 140, 3

\bibitem[Morton(1978)]{mor78} Morton, D. C. 1978, \apj, 222, 863

\bibitem[Morton(2003)]{mor03} Morton, D. C. 2003, \apjs, 149, 205

\bibitem[Nollett \& Burles(2000)]{nol00} Nollett, K. M., \& Burles, S.
2000, \prd, 61, 123505

\bibitem[Oliveira et al.(2005)]{oli05} Oliveira, C. M., Dupuis, J., Chayer, 
P., \& Moos, H. W. 2005, \apj, 625, 232

\bibitem[Oliveira et al.(2003)]{oli03}
Oliveira, C. M., H\'{e}brard, G., Howk, J. C., Kruk, J. W., Chayer, P., \&
Moos, H. W. 2003, ApJ, 587, 235

\bibitem[Oliveira et al.(2006)]{oli06} Oliveira, C. M., Moos, H. W., 
Chayer, P., \& Kruk, J. W. 2006, \apj, accepted (astro-ph/0601114).

\bibitem[Parise et al.(2002)]{par02} Parise, B., et al. 2002, \aap, 393, L49

\bibitem[Parise et al.(2004)]{par04} Parise, B., Castets, A., Herbst, E., 
Caux, E., Ceccarelli, C., Mukhopadhyay, L., \& Tielens, A.~G.~G.~M. 2004,
\aap, 416, 159  

\bibitem[Peeters et al.(2004)]{pee04} Peeters, E., Allamandola, L. J., 
Bauschlicher Jr., C. W., Hudgins, D. M., Sandford, S. A., \& 
Tielens, A. G. G. M. 2004, \apj, 604, 252

\bibitem[Pendleton \& Allamandola(2002)]{pen02} Pendleton, Y. J., \&
Allamandola, L. J. 2002, \apjs, 138, 75

\bibitem[Piskunov et al.(1997)]{pis97}
Piskunov, N., Wood, B. E., Linsky, J. L., Dempsey, R. C., \& Ayres, T. R.
  1997, ApJ, 474, 315

\bibitem[Prochaska, Tripp, \& Howk(2005)]{pro05} Prochaska, J. X., Tripp, 
T. M., \& Howk, J. C. 2005, \apj, 620, L39

\bibitem[Redfield \& Linsky(2000)]{red00} Redfield, S., \& Linsky, J. L. 2000, 
\apj, 534, 825

\bibitem[Redfield \& Linsky(2002)]{red02} Redfield, S., \& Linsky, J. L. 2002, 
\apjs, 139, 439

\bibitem[Redfield \& Linsky(2004a)]{red04a} Redfield, S., \& Linsky, J. L. 
2004a, \apj, 602, 776

\bibitem[Redfield \& Linsky(2004b)]{red04b} Redfield, S., \& Linsky, J. L. 
2004b, \apj, 613, 1004

\bibitem[Reynoso \& Dubner(1997)]{rey97} Reynoso, E. M., \& Dubner, G. M. 1997,
A\&AS, 123, 31

\bibitem[Rogers et al.(2005)]{rog05} Rogers, A. E. E., Dudevoir, K. A., 
Carter, J. C., Fanous, B. J., Kratzenberg, E., \& Bania, T. M. 2005,
\apj, 630, L41 

\bibitem[Romano et al.(2003)]{rom03} Romano, D., Tosi, M., Matteucci, F., \& 
Chiappini, C. 2003, \mnras, 346, 295

\bibitem[Romano et al.(2006)]{rom06} Romano, D., Tosi, M., Chiappini, C., \&
Matteucci, F. 2006, \mnras, accepted ~(astro-ph/0603190)

\bibitem[Sahnow et al.(2000)]{sah00} Sahnow, D. J. et al. 2000, \apj, 538, L7

\bibitem[S\'anchez et al.(2006)]{san06} S\'anchez, A. G., Baugh, C. M., 
Percival, W. J., Peacock, J. A., Padilla, N. D., Cole, S., Frenk, C. S.,
Norberg, P. 2006, \mnras, 366, 189

\bibitem[Savage \& Sembach(1996)]{sav96} Savage, B. D. \& Sembach, K. R. 1996,
\araa, 34, 279

\bibitem[Savage et al.(1977)]{sav77} Savage, B. D., Drake, J. F., Bohlin, 
R. C., \& Budich, W. 1977, \apj, 216, 291

\bibitem[Seaton \& Badnell(2004)]{sea04} Seaton, M. J. \& Badnell, N. R. 2004,
MNRAS, 354, 457

\bibitem[Sembach \& Savage(1996)]{sem96} Sembach, K. R., \& Savage, B. D. 
1996, \apj, 457, 211

\bibitem[Sembach et al.(2004)]{sem04} Sembach, K. R., et al. 2004, \apjs, 150, 
387

\bibitem[Sfeir et al.(1999)]{sfe99} Sfeir, D. M., Lallement, R., Crifo, F., \& 
Welsh, B. Y. 1999, \aap, 346, 785

\bibitem[Shull et al.(2005)]{shu05} Shull, J. M., Anderson, K. L., 
Tumlinson, J., et al. 2005, \apj, submitted

\bibitem[Shull \& York(1977)]{shu77} Shull, J. M. \& York, D. G. 1977, \apj,
211, 803


\bibitem[Sofia, Cardelli, \& Savage(1994)]{sof94} Sofia, U. J., 
Cardelli, J. A., \& Savage, B. D. 1994, \apj, 430, 650

\bibitem[Sonneborn et al.(2000)]{son00}
Sonneborn, G., Tripp, T. M., Ferlet, R., Jenkins, E. B., Sofia, U. J.,
  Vidal-Madjar, A., \& Wozniak, P. R. 2000, ApJ, 545, 277

\bibitem[Sonneborn et al.(2002)]{son02}
Sonneborn, G., et al. 2002, ApJS, 140, 51

\bibitem[Spergel et al.(2003)]{spe03} Spergel, D. N., et al. 2003, \apjs, 148, 
175

\bibitem[Spitzer, Cochran, \& Hirshfeld(1974)]{spi74} Spitzer, L. Jr., Cochran,
W. D., \& Hirshfeld, A. 1974, \apjs, 28, 373

\bibitem[Spitzer \& Fitzpatrick(1993)]{spi93} Spitzer, L. Jr., \& Fitzpatrick, 
E. L. 1993, \apj, 409, 299

\bibitem[Spitzer \& Jenkins(1975)]{spi75} Spitzer, L. Jr. \& Jenkins, E. B.
1975, \araa, 13, 133


\bibitem[Tielens(1983)]{tie83} Tielens, A. G. G. M. 1983, \aap, 119, 177

\bibitem[Turner(1990)]{tur90} Turner, B. E. 1990, \apj, 362, L29

\bibitem[van der Tak et al.(2002)]{tak02} van der Tak, F.~F.~S., Schilke, P.,
M\"uller, H.~S.~P., Lis, D.~C., Phillips, T.~G., Gerin, M., \& Roueff, E.
2002, \aap, 388, L53 

\bibitem[van Steenberg \& Shull(1988)]{van88} van Steenberg, M. E. \&
Shull, J. M. 1988, \apjs, 67, 225

\bibitem[Vennes et al.(2000)]{ven00}
Vennes, S., Polomski, E. F., Lanz, T., Thorstensen, J. R., Chayer, P., \&
  Gull, T. R. 2000, ApJ, 544, 423

\bibitem[Vidal-Madjar et  al.(1977)]{vid77} Vidal-Madjar, A., Laurent, C., 
Bonnett, R. M., \& York, D. G. 1977, \apj, 211, 91

\bibitem[Wakker \& Mathis(2000)]{wak00} Wakker, B. P., \& Mathis, J. S. 2000,
\apj, 544, L107

\bibitem[Welty, Hobbs \& Kulkarni(1994)]{wel94} Welty, D. E., Hobbs, L. M.,
\& Kulkarni, V. P. 1994, \apj, 436, 152

\bibitem[Williger et al.(2005)]{wil05} Williger, G. M., Oliveira, C., 
H\'ebrard, G., Dupuis, J., Dreizler, S., \& Moos, H. W.  2005, \apj, 
625, 210

\bibitem[Wilson \& Rood(1994)]{wil94} Wilson, T. L. \& Rood, R. T. 1994, 
\araa, 32, 191

\bibitem[Wood, Alexander, \& Linsky(1996)]{woo96} 
Wood, B. E., Alexander, W. R., \& Linsky, J. L. 1996,
ApJ, 470, 1157 

\bibitem[Wood, Linsky, \& Zank(2000)]{woo00} 
Wood, B. E., Linsky, J. L., \& Zank, G. P. 2000, ApJ, 537, 304 

\bibitem[Wood et al.(2002)]{woo02} Wood, B. E., Linsky, J. L., H\'ebrard, G., 
Vidal-Madjar, A., Lemoine, M., Moos, H. W., Sembach, K. R., \& Jenkins, E. B. 
2002, \apjs, 140, 91

\bibitem[Wood et al.(2004)]{woo04} Wood, B. E., Linsky, J. L., H\'ebrard, G., 
Williger, G. M., Moos, H. W., \& Blair, W. P. 2004, \apj, 609, 838

\bibitem[Yan et al.(2004)]{yan04} Yan, H., Lazarian, A., \& Draine, B. T. 2004,
\apj, 616, 895

\bibitem[York(1983)]{yor83}
York, D. G. 1983, ApJ, 264, 172

\bibitem[York \& Rogerson(1976)]{yor76} York, D. G. \& Rogerson, J. B. 1976,
\apj, 203, 378 

 \end{thebibliography}
\end{document}